\RequirePackage{tikz}
\documentclass[]{interact}

\usepackage[numbers,sort&compress]{natbib}
\bibpunct[, ]{[}{]}{,}{n}{,}{,}
\makeatletter
\def\NAT@def@citea{\def\@citea{\NAT@separator}}
\makeatother

\usepackage{hyperref}
\usepackage{mathtools}
\usepackage{subcaption}
\usepackage{breqn}
\usepackage{algorithm}
\usepackage{algpseudocode}
\usepackage{microtype}

\newcommand{\specialcell}[2][c]{%
  \begin{tabular}[#1]{@{}c@{}}#2\end{tabular}}

\usepackage{xcolor}
\definecolor{darkblue}{HTML}{785EF0}
\definecolor{blue}{HTML}{739AFF}
\definecolor{mag}{HTML}{DC267F}
\definecolor{yellow}{HTML}{FFB000}
\tikzstyle{inputdynsys}=[ellipse, draw=black, very thick, inner sep = 0pt, minimum height = 10mm, minimum width= 30mm]
\tikzstyle{dynsysnode} = [rectangle, draw=black, very thick, inner sep = 0pt, minimum height = 5mm, minimum width = 15mm]
\tikzstyle{wv} = [rectangle, draw=black, very thick, inner sep=0pt, minimum height=2mm, minimum width=15mm]
\tikzstyle{outputdynsys} = [circle, draw=black, very thick, inner sep = 0pt, minimum size = 20mm]
\tikzstyle{falseoutput} = [circle, draw=black, thin, inner sep = 0pt, minimum size = 20mm]

\tikzstyle{inputnode}=[circle, draw=black, fill=mag, very thick, inner sep = 0pt, minimum size=10mm]
\tikzstyle{innernode}=[circle, draw=black, fill=blue, very thick, inner sep = 0pt, minimum size=10mm]
\tikzstyle{outputnode}=[circle, draw=black, fill=yellow, very thick, inner sep = 0pt, minimum size=10mm]

\tikzstyle{inputset}=[rectangle,draw=black, fill=mag, very thick, inner sep = 0pt, minimum size=10mm]
\tikzstyle{innerset}=[rectangle,draw=black, fill=blue, very thick, inner sep = 0pt, minimum size=10mm]
\tikzstyle{outputset}=[rectangle,draw=black,  fill=yellow, very thick, inner sep = 0pt, minimum size=10mm]

\tikzstyle{inputsmall}=[rectangle, draw=black, fill=red, very thick, inner sep = 0pt, minimum size=7mm]
\tikzstyle{innersmall}=[circle, draw=black, fill=cyan, very thick, inner sep = 0pt, minimum size=7mm]
\tikzstyle{outputsmall}=[rectangle, draw=black, fill=teal, very thick, inner sep = 0pt, minimum size=7mm]

\begin{document}

\title{Reservoir Computing Benchmarks:
a tutorial review and critique
}

\author{
\name{Chester Wringe\textsuperscript{a},
Martin Trefzer\textsuperscript{b}
and 
Susan Stepney\textsuperscript{a}\thanks{CONTACT Susan Stepney. Email: susan.stepney@york.ac.uk  ORCID: 0000-0003-3146-5401}
}
\affil{
\textsuperscript{a}Department of Computer Science, University of York, YO10~5DD, UK;\\
\textsuperscript{b}School of Physics, Engineering and Technology, University of York, YO10~5DD, UK
}
}

\maketitle

\begin{abstract}Reservoir Computing is an Unconventional Computation model to perform computation on various different substrates, such as recurrent neural networks or physical materials. The method takes a `black-box' approach, training only the outputs of the system it is built on. As such, evaluating the computational capacity of these systems can be challenging. 
We review and critique the evaluation methods used in the field of reservoir computing.
We introduce a categorisation of benchmark tasks.
We review multiple examples of benchmarks from the literature as applied to reservoir computing, and note their strengths and shortcomings.
We suggest ways in which benchmarks and their uses may be improved to the benefit of the reservoir computing community. 
\end{abstract}


\section{Introduction}\label{intro}

When evaluating a given system, a common approach is to use benchmarks. 
Many authors present their results using benchmark tasks, often using an appeal to the literature to justify the use of any given benchmark.

Reservoir Computing is a form of machine learning particularly suitable for addressing time series and other temporal processing tasks.

There are two main variants. The Echo State Network (ESN) was first introduced by Jaeger~\citep{Jaeger_undated-ek,Jaeger_2010,Jaeger2007-ar}, as a way of training Recurrent Neural Networks efficiently. A similar model, the Liquid State Machine (LSM), based on a spiking network model was later introduced by Maass~\citep{Maass2002-rq}, and was intended to present a more biologically plausible way of training Neural Networks.

Reservoir Computers are uniquely suited to temporal tasks, which means that most of the benchmarks used to evaluate them are of a similar temporal nature. For historical reasons pertaining to their origin in neural networks, Reservoir Computing is also frequently used in classification tasks. 
Little is written on what these benchmarks are, why they are used, how to use them, what makes them a useful measure of the performance of a Reservoir Computer, or what different benchmarks have in common.
Classification is a less intuitive way of using Reservoir Computing, and the tasks typically require some processing to be adapted to suit Reservoir Computing. For this reason, we  focus primarily on temporal benchmarks. 

This review aims to answer some of these questions, 
offers a general paradigm through which to view benchmarks, 
and proposes some best practices when using benchmarks. 

\subsection{What Are Benchmarks?}\label{what-are-benchmarks}

We introduce some terminology.
Any given thing that one uses a Reservoir Computer to do is a \textit{task}.
A \textit{benchmark} is a task used to evaluate the Reservoir Computer itself. 
This is as opposed to a \textit{problem}, a task where a Reservoir Computer is used to provide a unique, insightful, or specific answer. 
While any given task may sometimes be employed as a benchmark (evaluating the RC), and sometimes as a problem (using the RC), 
we believe that the distinction between them is helpful to clarify arguments being made.

Benchmarks and problems can be distinguished by examining what the author argues. 
If the work is about solving a task, 
and the Reservoir Computer merely a means to an end, 
then the task in question is a problem. 
Conversely, if the work is about Reservoir Computers, 
or a specific implementation of Reservoir Computing, 
and the task is used to illustrate or evaluate the argument posed, 
then the task in question is a benchmark. 
A problem is exploratory, a benchmark is comparative. 
Whether the comparison is to other works in the literature, 
other implementations produced in the work, 
or even an arbitrary standard with no other basis for existing, 
the comparative element remains there. 

Frequent recurrence of a task in the literature is neither a necessary nor a sufficient condition for it being a benchmark: certain problems may recur because Reservoir Computing is well-suited to solving them; certain tasks may be used as benchmarks for specific reasons, or for the first time.
However, if a problem is frequently approached through the lens of reservoir computing, it may become a benchmark; see, for example, the spoken digits benchmark (sec.\ref{spoken-digits}).

\subsection{Why Are Benchmarks Used?}

Given benchmarks are tasks used in a specific way,
are they indeed a good way of evaluating Reservoir Computing in general,
or just for those specific tasks? 
This is not a commentary on the quality or effectiveness of individual benchmarks (although such commentary is made below) but a look at the purpose of their use. 
Another way of evaluating Reservoir Computers is CHARC~\citep{Dale2019-rm}, and there are other methods for evaluating generic task-independent properties such as Kernel Rank, Generalisation Rank, and Memory Capacity (sec.\ref{sec:property-measures}). So why use problem-based benchmarks, specific to a single task, instead of the task-independent property measures like CHARC?

We propose that the answer is twofold: a benchmark can help place a specific instance of a Reservoir Computer in context with the literature, and it is a measure that is both practical and quantifiable.

Placing our work in the context of the literature is useful for several reasons: we can compare actual results, often (although not always) through the use of a common and well-described experimental procedure. One such example is the procedure of the spoken digits task, as first described in~\citep{Verstraeten2005-fk}. It can also help us showcase particular properties of a Reservoir Computer, as do the long-short term memory benchmarks~\citep{Jaeger2012-ze}.

The quantifiable aspect is also useful: by being able to express the performance of a Reservoir Computer on a given benchmark, one can say that it is better or worse than others, as opposed to simply `different'. 

Although several good reasons for using benchmarks exist, care should be taken when choosing specific benchmarks for different cases.
Certain benchmarks may be better suited to a given argument than others, 
while other arguments may be better served by using something else entirely. 

\subsection{Types of Reservoir Computing works}

In the literature, benchmarks can be used for practical reasons (the benchmark is especially suited to make the argument the work wishes to), or historical ones (other works of similar sort use that benchmark).

We divide the Reservoir Computing literature into three categories, similar to those described by \cite{Tanaka2019-sh}. Although not every reference falls cleanly into one or other of these, the categorisation provides a useful tool to identify what arguments a work is making, and which benchmarks may be best suited to support those arguments. 
We propose the following categories:

 \paragraph*{The Reservoir Computing approach to problem X.}
    These works typically focus on particular problems, and whether Reservoir Computing can provide a solution. An example is investigating the usefulness of Reservoir Computing when controlling robots~\citep{Ploger2003-ex}. 

    Works in this category typically do not use benchmarks, but they may still have a comparative element to them, such as comparing the Reservoir Computer’s performance to the current state of the art; for example, such as discrete Hidden Markov Models in investigations of speech recognition~\citep{Verstraeten2005-fk,F_Triefenbach_A_Jalal_B_Schrauwen_and_J-P_Martens2010-cq}.
    
    Although this category does not use benchmarks in service of the main argument,
     it is helpful to distinguish it, to provide a contrast to the categories that do, described below.
    It can also help to identify tasks that Reservoir Computing may be well suited to, such as Spoken Digit Recognition~\citep{Verstraeten2005-fk}, or predicting spatio-temporally chaotic systems like the Kuramoto--Sivashinsky Equation~\citep{Kuramoto1978-up},
    that may subsequently become good benchmarks.

\paragraph*{A Novel or Improved Reservoir Computing model.}
    Works in this category typically build on the original ESN~\citep{Jaeger_undated-ek} and LSM~\citep{Maass2002-rq} Reservoir Computing Models, in order to change or improve them. They may use benchmarks to demonstrate that their model is better than previous models; for example, where Jaeger~\citep{Jaeger2007-mb} introduces the concept of Leaky Integrator based ESNs. They may also use them to show that their model performs similarly to other models, but has other advantages, such as being simpler to implement in hardware. Examples of these include the introduction of the delay-line architecture~\citep{Appeltant2011-el}, and in the analysis of reservoir topologies~\citep{Rodan2010-ku}.
    
\paragraph*{A physical implementation of a Reservoir Computer.} 
    These works are typically investigations into whether a particular material substrate is suitable as an implementation of a hardware Reservoir Computer, in terms of both performance and practicality \cite{Stepney-2024-NACO}. One of the first examples in this category, instantiating an LSM in a bucket of water~\citep{Fernando2003-oo}, was used to illustrate properties of Reservoir Computers. The works in this category are reviewed by~\citep{Tanaka2019-sh}; examples include a reservoir instantiated in a swarm model~\citep{Lymburn2021-ef}, in a robot modelled on an octopus arm~\citep{Nakajima2013-lx}, and in carbon nanotubes~\citep{Dale2016-jc}.

\subsection{Structure of this review}\label{structure-of-this-paper}

Other authors provide overviews of a variety of RC benchmarks and problems, and some categorisation of tasks. 
For example, \cite{Yan2024} provides a compact categorisation of many benchmarks and applications.
Here, we take many of the more well-know, and well-used, benchmarks,
delve into their details, variations, and limitations,
and critique their use and applicability.

In section \ref{sec:prelim} we provided some preliminary definitions and classifications: dynamical systems terminology, input and output types, 
supervised and unsupervised learning, and reservoir performance evaluation measures.
In section \ref{sec:tasks} we identify the major classes of RC benchmarks reviewed here, and explain how they differ.
We next review multiple examples of each of these classes:
imitation tasks (section \ref{sec:imitate}),
prediction tasks (section \ref{sec:predict}),
computation tasks (section \ref{comp-tasks}),
classification tasks (section \ref{classification-tasks}),
and property measures (section \ref{sec:property-measures}).
The specific benchmarks we review in these sections are given in table \ref{table:dyn-sys-task}. 
We finish by describing best practices in choosing, using, and comparing benchmarks for reservoir computing, in section~\ref{best practice}.

\begin{table*}[tp]
\begin{center}\small
\begin{tabular}{lllc}
\toprule%
task type & dynamics & benchmark & section \\
\midrule
imitation & known & \\ 

&& NARMA-$N$ & \ref{subsubsec:narma} \\
imitation & unknown & \\

&& channel equalisation & \ref{subsubsec:channel-equalisation} \\
&& two-jointed arm & \ref{sec:2jointarm}\\
&& Van del Pol controller & \ref{sec:van-der-pol} \\
&& pole balancing & \ref{sec:pole-bal} \\

prediction & known & \\
&& Mackey--Glass equation & \ref{sec:Mackey-Glass}\\
&&Multiple Superimposed Oscillators (MSO) & \ref{sec:mso} \\
&&lazy figure eight  & \ref{sec:lazy8} \\
&& Lorenz chaos & \ref{sec:lorenz}\\
&& Kuramoto--Sivashinski equation  & \ref{sec:ks-eqn} \\

 & unknown & \\
&& sunspot numbers  & \ref{sec:sunspots} \\
&&  Santa Fe laser data & \ref{sec:Santa-Fe} \\
&& McMaster IPIX radar data & \ref{sec:ipix radar} \\

computation\\
&& XOR  & \ref{sec:xor} \\
&&  parity  & \ref{sec:parity}\\

classification\\
&& MNIST handwritten digits & \ref{sec:mnist}\\
&& spoken digits & \ref{spoken-digits} \\
&& Japanese vowels & \ref{sec:vowel} \\
&& Santa Fe sleep apnea data & \ref{sec:Santa-Fe-apnea} \\

property\\
&& linear memory capacity (MC) & \ref{sec:lmc}\\
&&  nonlinear memory capacity & \ref{sec:IPC}\\
&& kernel rank (KR)  & \ref{sec:KR}\\
&&  generalisation rank (GR) & \ref{sec:GR}\\

\bottomrule
\end{tabular}
\caption{
The benchmarks reviewed here, arranged by kind of task.
}\label{table:dyn-sys-task}
\end{center}
\end{table*}

\section{Reservoir computing terminology}\label{sec:prelim}

Here we introduce and define some terminology and notation used in our reservoir computing benchmark review. 
\subsection{Dynamical Systems}\label{dynamical-systems}
A dynamical system is described by its state's time evolution, within a given state space. 
A continuous time dynamical system can be defined with an ordinary differential equation, as 
$\dot{\mathbf{x}} = f(\mathbf{x}(t), \mathbf{u}(t), \mathbf{b})$,
where $t \in \mathbb{R}$ is the time, $\mathbf{x}$ is the state, $\mathbf{u}$ is the input,
and $\mathbf{b}$ is a time independent parameter.
A discrete time dynamical system can be defined with a difference equation, as
$\mathbf{x}(t+1) = f(\mathbf{x}(t), \mathbf{u}(t), \mathbf{b})$,
where $t \in \mathbb{N}$ is the time.
More sophisticated dynamical systems can be defined; for example,
using delay differential equations to incorporate memory of previous states,
using partial differential equations to incorporate spatial properties.

The term $\mathbf{u}(t)$ captures the inputs to
an \textit{open} (non-autonomous) dynamical system.
If $\mathbf{u}(t) = 0$,
the system has no inputs, and is 
\textit{closed} (autonomous),
with its behaviour depending on only its initial state and other fixed parameters. 
A given system's internal dynamics $f$ and inputs $\mathbf{u}$ may be understood and known, or may be unknown. 

\subsection{Echo State Networks}

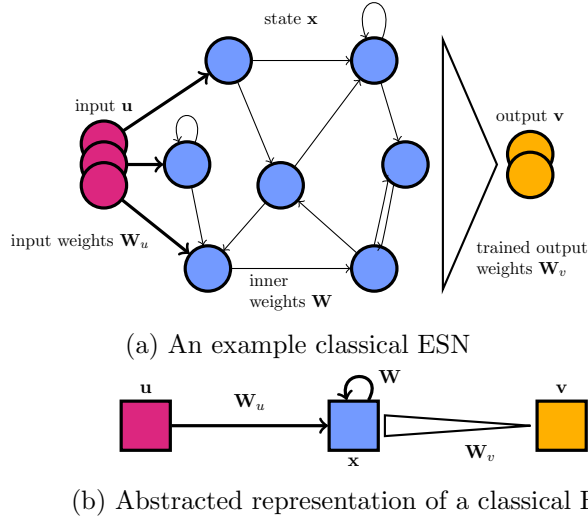
\begin{figure}[tp]
\mbox{
\begin{subfigure}[b]{0.9\linewidth}
\centering
\begin{tikzpicture}[scale = 0.55, every node/.style={scale=0.6}]
\node at (-5, 0.5)[inputnode, label=input $\mathbf{u}$] (input) {};
\node at (-5, 0) [inputnode] (input1) {};
\node at (-5, -0.5) [inputnode] (input2) {};
\node at (-2, 2.5)[innernode] (1) {};
\node at (-3, 0)[innernode] (2) {};
\node at (-2.5, -2.5)[innernode] (3) {};
\node at (-0.75, -0.5)[innernode] (4){};
\node at (1.5, -2.5)[innernode] (5) {};
\node at (1.5, 2.5)[innernode] (6) {};
\node at (2.25, 0)[innernode] (7) {};
\node at (5.25, 0.25) [outputnode, label=output $\mathbf{v}$] (output) {};
\node at (5.25, -0.25) [outputnode] (output1) {};

\node at (-0.5, 3.5) [] (inner) {state $\mathbf{x}$};

\draw [->][very thick] (input) -- (1);
\draw[->][very thick] (input1) -- (2);
\draw[->][very thick] (input2) -- node[below left] {input weights $\mathbf{W}_u$} (3);

\draw[black, thick] (3.15, -3) -- node[right, yshift=-20pt, align=left] {trained output\\weights $\mathbf{W}_v$}(4.45, 0) -- (3.15, 3) -- cycle;

\draw[->][thin] (1) -- (4);
\draw[->][thin] (2) to [out=70,in=110,looseness=6] (2);
\draw[->][thin] (2) -- (3);
\draw[->][thin] (4) -- (3);
\draw[->][thin] (3) -- node[below, align=left] {inner\\weights $\mathbf{W}$} (5);
\draw[->][thin] (5) -- (4);
\draw[->][thin] (5) to [out=90, in=-135, looseness=-6] (7);
\draw[->][thin] (7) to [out=-120, in=70, looseness=-6](5);
\draw[->][thin] (4) -- (6);
\draw[->][thin] (1) -- (6);
\draw[->][thin] (6) to [out=70,in=110,looseness=8] (6);
\draw[->][thin] (6) -- (7);
\end{tikzpicture}
\caption{An example classical ESN}\label{classical-ESN}
\end{subfigure}   
}
\mbox{
\begin{subfigure}[b]{\linewidth}
\centering
\begin{tikzpicture}[scale = 0.55, every node/.style={scale=0.65}]
\node at (0, 4) [inputset, label=$\mathbf{u}$] (input) {};
\node at (5, 4) [innerset, label=below: $\mathbf{x}$] (inner) {};
\node at (10, 4) [outputset, label=$\mathbf{v}$] (output) {};

\draw[->][very thick] (input) -- node[above, yshift=5pt, align=left] {$\mathbf{W}_u$} (inner);
\draw[->][very thick] (inner) to [out=60, in=105, looseness=4] node[right, xshift=6pt, align=left] {$\mathbf{W}$} (inner);

\draw[black, thick] (5.75, 4.25) -- node[below right, yshift=-10pt, align=left] {$\mathbf{W}_v$}(9.25, 4) -- (5.75, 3.75) -- cycle;
\end{tikzpicture}
\caption{Abstracted representation of a classical ESN}\label{classical-ESN-elements}
\end{subfigure}
}
\caption[The classical ESN model]{\small
(a) An example classical ESN with $7$ nodes. This ESN takes a 3-d vector of inputs $\mathbf{u}$, which are sent to the inner state $\mathbf{x}$ through weighted edges $\mathbf{W}_u$. The weights within the inner state, $\textbf{W}$, are recurrent and randomly set. The 2-d output vector $\mathbf{v}$ receives the inner state through trained edges $\mathbf{W}_v$.
(b) An abstract representation of the different components of a general ESN.}\label{classical-ESN-full}
\end{figure}

Echo State Networks (fig. \ref{classical-ESN-full}) are modelled as random recurrent neural networks,
which have the form of a discrete time dynamical system.
The minimal form of an ESN is:
\begin{eqnarray}
    \mathbf{x}(t+1) &=& f(\mathbf{W}\,\mathbf{x}(t) + \mathbf{W}_u \mathbf{u}(t)) \label{eqn:ESN}
    \\
    \mathbf{v}(t+1) &=& \mathbf{W}_v\,\mathbf{x}(t) \label{eqn:ESN-out}
\end{eqnarray}
Equation \ref{eqn:ESN}
defines the dynamics of the reservoir, where
$\mathbf{x}$ is the $N$-d reservoir state; 
$\mathbf{W}$ is a random $N\times N$ weight matrix; 
$\mathbf{u}$ is the $N_u$-d input;
$\mathbf{W}_u$ is a random input $N\times N_u$ weight matrix;
and $f$ is a non-linear function (typically $\tanh(.)$ or other sigmoid function).
Further terms may be included to allow leakiness, feedback, and so on.

Equation \ref{eqn:ESN-out}
defines the $N_v$-d output $\mathbf{v}$ as an observation of the reservoir state through a trained output $N_v\times N$ weight matrix $\mathbf{W}_v$.\footnote{%
Some authors use $\mathbf{u}(t+1)$ and/or $\mathbf{v}(t)$ in their defining equations.
See~\citep{Stepney:2021-UCNC} for a discussion.
}

An ESN is typically trained to match a target output $\hat{\mathbf{v}}$, often using ridge regression; a guide to this as well as a discussion of how to set various RC parameters can be found in~\cite{Lukosevicius2012}.

Some other Reservoir Computing paradigms are:
\begin{itemize}
    \item \textit{In materio} reservoirs~\citep{Tanaka2019-sh,Stepney-2024-NACO} are a form of physical RC where the recurrent neural network model is (approximately) represented using some physical non-linear dynamical system. 
    \item \textit{Liquid State Machines}~\citep{Fernando2003-oo} are a Reservoir Computing paradigm based on Spiking Neural Networks.
    \item \textit{Delay-line} reservoir~\citep{Appeltant2011-el} uses a single dynamical node, using a mask on the input such that it is multiplexed in time, rather than in space.
    These are commonly instantiated as a form of physical RC, but may also be simulated digitally.
\end{itemize}

\subsection{Reservoir Input}
\label{sec:reservoir-input}

\subsubsection{Dimensionality}

The general reservoir equation (eqn.~\ref{eqn:ESN}) allows for a vector-valued input at each timestep.
Most of the benchmarks described here, such as NARMA (section \ref{subsubsec:narma}) 
and the sunspot prediction benchmark (section \ref{sec:sunspots}), have a single-valued, or scalar-valued, input at each timestep
($N_u = 1$). 
Some more complex multi-valued, or vector-valued, ($N_u > 1$) input benchmarks, such as the XOR task \citep{Fernando2003-oo}, the Van der Pol Oscillator task \citep{10191630},
and the sleep apnea task (section \ref{sec:Santa-Fe-apnea})
are also available.


\subsubsection{Stationary v. non-stationary input}

The input to a reservoir is \textit{stationary} in the statistical sense if its statistical properties are not a function of time:
it has no long term systematic changes or trends, no preferred zero time point.
So a random uniform input is stationary;
a steadily growing value is not.
A cyclic, or seasonal, data set (for example, the sunspot data set in section \ref{sec:sunspots}) may be stationary if the statistical properties are calculated over timescales longer than the cycle length, but not over shorter ones.

A stationary data set is necessarily infinite (in theory), since a beginning or an end are special time points.
In practice, no data set is infinite, but provided it is long enough that its end points are not visible to the system under test, and its statistical properties are the same across the data set, that is sufficient.

A data set generated from an equation can be stationary,
and has the advantage that the data set can be extended, simply by running the generating equation for longer.
Gilpin \cite{DBLP:journals/corr/abs-2110-05266}
provides a catalogue of 131 chaotic dynamics systems equations,
suitable for generating stationary time series for benchmarks.
The catalogue includes the Mackey--Glass (section \ref{sec:Mackey-Glass}) and Lorenz systems (section \ref{sec:lorenz}).

Experimentally-generated data sets can be stationary
in the limited sense described above,
if they can in principle be extended, by running the experiment for longer, and taking further observations.
The Santa Fe laser data (section \ref{sec:Santa-Fe}) and sleep apnea data (section \ref{sec:Santa-Fe-apnea}) are such examples, provided that the statistical values are calculated over long enough timescales to encompass the semi-periodic variations; the sunspot data (section \ref{sec:sunspots}) is also (we hope!) such an example.

Many experimental data sets are, however, naturally bounded,
and so non-stationary.
These include various spoken word (section \ref{spoken-digits}) and spoken vowel (section \ref{sec:vowel}) data sets: each word or vowel sound is discrete, with a beginning and an end point.

\subsubsection{Washout data}\label{sec:washout}
Typically, an initial subset of the data is used for a \textit{washout} period.
This puts the reservoir into a state 
so that results are independent of its earlier history, such as running a previous experiment, or of being switched off. 
This forgetting is possible because of the reservoir's fading memory property.

The need for a washout period can limit the training and testing periods of experimental data sets.
Even for data sets that are effectively stationary, they nevertheless have a limited number of data points.

There is little guidance in the literature on how washout is handled for non-stationary data (for example, spoken digits, section~\ref{spoken-digits}).  Several approaches are possible, depending on the precise application:
(i) provide explicitly zero input as washout, and let the reservoir settle into a resting state;
(ii) present the data multiple times;
(iii) use an initial subsequence of the data as washout.

\subsection{Reservoir Output}\label{range-and-dedicated-output}

\subsubsection{Real valued encoding}
Here, the reservoir has one or more output nodes, each of which takes a real value, from which the result is read and used as the output of the task. 
If the output is a scalar, the reservoir has a single node encoding the scalar value, otherwise the reservoir has one node for each dimension of the output vector~$\textbf{v}$ (see equation~\ref{eqn:ESN-out}).

The output is typically scaled to some specific range, such as $[-1,1]$.
The precision of the output is constrained by encoding or readout methods. 

This encoding method is typically used for tasks where the reservoir is trained to behave like some dynamical system with a non-binary (or even continuous) encoding.  The time-series input is matched by a corresponding time-series output of real values.

\subsubsection{Categorical encoding}\label{sec:class-cat-out}
When the reservoir's output is one of several categories, such as in classification tasks, a different form of output encoding can be used.

\paragraph*{Multiple categories: one-hot encoding.}
In `one-hot' encoding, the reservoir has $N$ output nodes, one for each categorical value. At any given timestep, the output of the reservoir is taken to be the value of the node with the highest response (the one `hot' node). As such, this encoding is suited to many classification tasks. 

In this encoding, each node is trained on a binary choice : whether the output of the reservoir is or is not its corresponding value. 
When using this encoding, it is possible that multiple output nodes to have non-zero values.
This may count as an error.
In other cases, the strongest response be chosen. 
Intermediate values may correspond to uncertainty of ambiguity measures.

\paragraph*{Two categories: a single binary choice.}
When there are only two categories, or a single binary choice,
the number of nodes can be reduced to one,
with `0' denoting one category, and `1' denoting the other.
In this case, distance from the 0 or 1 results may represent categorisation error.

\section{A taxonomy of tasks}\label{sec:tasks}

\subsection{Main classes of benchmark tasks}
We identify five main classes of benchmark task.
The first two relate to training a reservoir to \textit{imitate} an open (non-autonomous) dynamical system, or to \textit{predict} a closed (autonomous) one.
The next two classes relate to reservoirs performing specific \textit{computations} on their inputs,
or \textit{classifying} their inputs.
The final class is a set of measures of specific \textit{properties} of reservoirs.
As the subject develops, and RCs are applied to wider classes of problems,
new classes of task may be devised.

\subsubsection{Imitation tasks}\label{sec:imitate-intro}

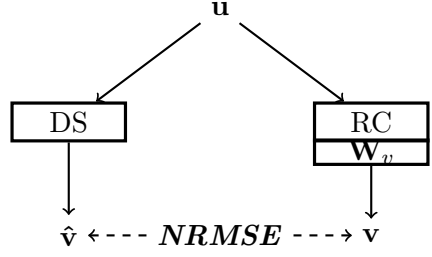
\begin{figure}[tp]
    \centering
    \begin{tikzpicture}
        \node at (0, 3) (u) {$\mathbf{u}$};
        \node at (-2, 1.5) [dynsysnode] (dynsys) {DS};
        \node at (2, 1.5) [dynsysnode] (res) {RC};
        \node at (2, 1.1) [wv] (wv) {$\mathbf{W}_v$};
        \node at (-2, 0) (vhat) {$\mathbf{\hat{v}}$};
        \node at (2, 0) (v) {$\mathbf{v}$};
        \node at (0, 0) (nrmse) {\textbf{\emph{NRMSE}}};

        \draw[->,  thick] (u) -- (dynsys);
        \draw[->,  thick] (u) -- (res);
        \draw[->,  thick] (dynsys) -- (vhat);
        \draw[->,  thick] (wv) -- (v);

        \draw[->, dashed,  thick] (nrmse) -- (vhat);
        \draw[->, dashed,  thick] (nrmse) -- (v);
    \end{tikzpicture}
    \caption{
    Training and testing stages of an imitation benchmark. 
    The reservoir RC is fed the same time series input  $\mathbf{u}$ as the target dynamical system DS. In the training stage, 
    the output weights $\mathbf{W}_v$ of the reservoir are trained such that the resulting reservoir output $\mathbf{v}$ resembles the target dynamical system output $\hat{\mathbf{v}}$, 
    the Normalised Root Mean Square Error (NRMSE, see section~\ref{sec:NRMSE}) can be used as an evaluation of the training. 
    In the testing stage, the trained reservoir output weights are used, 
    and we evaluate the reservoir using the error between the observed reservoir output $\mathbf{v}$ and the target dynamical system output  $\hat{\mathbf{v}}$.
    }
    \label{fig:imitation-diagram}
\end{figure}

In an imitation task (see Figure~\ref{fig:imitation-diagram}), the Reservoir Computer is trained to replicate the dynamics of an \textit{open} dynamical system, one with inputs. As such, it is given the same time series input as is given to the target dynamical system. 
This input may, for example, be uniform random noise, such as with NARMA (section~\ref{subsubsec:narma}),
or a regular frequency signal, such as used in the classic  sine wave generation task.
The reservoir is then trained to produce the same time series output that the target system produces. 

The performance of a Reservoir Computer on imitation tasks can be tested in two different ways. The first of these is to measure the error between the reservoir output values and the desired output values. In an imitation task, the desired output values are the output of the dynamical system being imitated. If the dynamical system being imitated is part of a greater whole, such as the control system of a robot, then another adequate test may be, `can the trained reservoir adequately replace the target system in situ'?


\subsubsection{Prediction tasks}\label{sec:prediction}

\begin{figure}[tp]
    \centering
    \begin{subfigure}{0.6\linewidth}
        \begin{tikzpicture}[scale = 0.9, every node/.style={scale=0.8}]
            \node at (0, 3) (t1) {$t$};
            \node at (0, 2) (t2) {$t+1$};
            \node at (0, 1) (t3) {$t+2$};

            \node at (1.5, 3.75) (dynsys) {Dynamical System};
            
            \node at (1.5, 3) [dynsysnode] (ds1) {};
            \node at (2.5, 3.25) (vhat1) {$~\mathbf{\hat{v}}(t)$};
            \node at (1.5, 2) [dynsysnode] (ds2) {};
            \node at (2.75, 2.25) (vhat2) {$~\mathbf{\hat{v}}(t+1)$};
            \node at (1.5, 1) [dynsysnode] (ds3) {};
            \node at (2.75, 1) (vhat3) {$~\mathbf{\hat{v}}(t+2)$};

            \node at (4.5, 3.75) (rc) {Reservoir Computer};
            \node at (4.5, 3) [dynsysnode] (r1) {};
            \node at (3.5, 2.75) (in1) {$\mathbf{u}(t)~$};
            \node at (4.5, 2) [dynsysnode] (r2) {};
            \node at (3.25, 1.75) (in2) {$\mathbf{u}(t+1)~$};


            \node at (7, 2) (out1) {$\mathbf{v}(t+1)$};
            \node at (7, 1) (out2) {$\mathbf{v}(t+2)$};

            \node at (4.9, 1) (nrmse) {\textbf{\emph{NRMSE}}};
            
            \draw[->,  thick] (ds1) -- (r1);
            \draw[->,  thick] (ds2) -- (r2);

            \draw[->,  thick] (r1) -- (out1);
            \draw[->,  thick] (r2) -- (out2);

            \draw[->, dashed,  thick] (nrmse) -- (vhat3);
            \draw[->, dashed,  thick] (nrmse) -- (out2);
        \end{tikzpicture}
    \caption{driven system}
    \end{subfigure}
    \begin{subfigure}{0.6\linewidth}
        \begin{tikzpicture}[scale = 0.9, every node/.style={scale=0.8}]
            \node at (-3, 3) (t1) {$t$};
            \node at (-3, 2) (t2) {$t+1$};
            \node at (-3, 1) (t3) {$t+2$};

            \node at (-1.5, 3.75) (dynsys) {Dynamical System};
            
            \node at (-1.5, 3) [dynsysnode] (ds1) {};
            \node at (-0.5, 3.25) (vhat1) {$~\mathbf{\hat{v}}(t)$};
            \node at (-1.5, 2) [dynsysnode] (ds2) {};
            \node at (-0.25, 2.25) (vhat2) {$~\mathbf{\hat{v}}(t+1)$};
            \node at (-1.5, 1) [dynsysnode] (ds3) {};
            \node at (-0.25, 1) (vhat3) {$~\mathbf{\hat{v}}(t+2)$};

            \node at (1.5, 3.75) (rc) {Reservoir Computer};
            \node at (1.5, 3) [dynsysnode] (r1) {};
            \node at (0.5, 2.75) (in1) {$\mathbf{u}(t)~$};
            \node at (1.5, 2) [dynsysnode] (r2) {};
            \node at (3, 1.75) (in2) {$\mathbf{u}(t+1)$};


            \node at (4, 2) (out1) {$\mathbf{v}(t+1)$};
            \node at (4, 1) (out2) {$\mathbf{v}(t+2)$};

            \node at (1.9, 1) (nrmse) {\textbf{\emph{NRMSE}}};
            
            \draw[->,  thick] (ds1) -- (r1);
            \draw[->,  thick] (ds2) -- (0, 2);

            \draw[->,  thick] (r1) -- (out1);
            \draw[->,  thick] (r2) -- (out2);
            \draw[->,  thick] (out1) -- (r2);

            \draw[->, dashed,  thick] (nrmse) -- (vhat3);
            \draw[->, dashed,  thick] (nrmse) -- (out2);
        \end{tikzpicture}
        \caption{free-running system}
    \end{subfigure}
    \caption{
    Training and testing stages of a prediction benchmark. During the training stage (a), the reservoir is given as inputs the target outputs of the dynamical system. The reservoir output weights are trained such that the reservoir outputs resemble the target outputs of the dynamical system. 
    There are two cases for testing.
    (a) \textit{Driven}: During the testing stage, the reservoir is again given the target outputs of the dynamical system. The reservoir outputs are compared to the  target outputs of the dynamical system.  
    (b) \textit{Free-running}:
    During the testing stage, the reservoir is fed back its own  output. The reservoir outputs are compared to the  target outputs of the dynamical system.
    }\label{fig:prediction-diagram}
\end{figure}
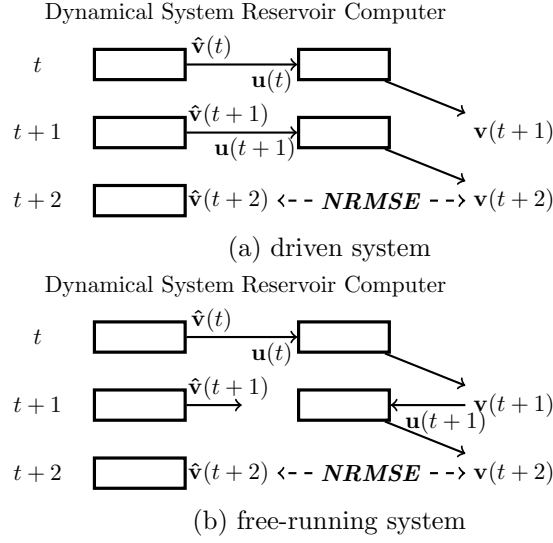

In a prediction task, the reservoir imitates a \textit{closed} dynamical system, one with no inputs. The reservoir must here predict the output of the dynamical system based on the previous output (figure \ref{fig:prediction-diagram}). 
This is useful for learning the behaviour of a dynamical system where the input is unknown, or perhaps does not even exist. However, that is not a requirement. Any dynamical system that can be used for an imitation task can also be used for a prediction task: the only difference is the information input to the reservoir.

During training, the reservoir is fed outputs from the target dynamical system as inputs, an the output weights trained to minimise the difference.
There are two options during the testing stage.
In the \textit{driven} case, the reservoir is driven with the target outputs as in the training stage.
This provides a one-step look-ahead prediction,
continually corrected by the actual system outputs to correct for drift.
In the \textit{free-running} case \citep{Jaeger2004-eu}
the reservoir is fed back its own previous output as its next input.
This provides a multi-step look-ahead prediction, potentially subject to accumulating errors.

\subsubsection{Computation tasks}

In a computation task, the focus is on the reservoir
performing specific logical or arithmetical computations on its inputs.
Examples of logical operations include XOR (section~\ref{sec:xor}), parity (section~\ref{sec:parity}), and logic gates \citep{Shougat2021-wd}.
Examples of arithmetical operations are covered by other classifications:
non-linear memory capacity (section~\ref{sec:IPC}) measures how well the reservoir can compute various polynomial functions of its inputs,
and imitating or predicting systems with known dynamics (section~\ref{sec:known}) requires the reservoir to compute those dynamics.

\subsubsection{Classification tasks}
Classification tasks have categorical outputs (section~\ref{sec:class-cat-out}): the task is to output the category corresponding to the input.  This reflects a typical use of neural networks, but here can explicitly capture the temporality of recognition, requiring memory over time.

There are three subclasses, depending on the form of the input to the reservoir
\begin{enumerate}
    \item static non-temporal input; single categorical output : example, MNIST (section~\ref{sec:mnist})
    \item non-stationary temporal input; single categorical output : example, spoken digits (section~\ref{spoken-digits}) and spoken vowels (section~\ref{sec:vowel})
    \item stationary temporal input; time-series categorical output : example, sleep apnea data (section~\ref{sec:Santa-Fe-apnea})
\end{enumerate}

Speech recognition is a traditional classification benchmark task \citep{George_R_Doddington_and_Thomas_B_Schalk1981-ix}. 
As such, it has also found a home as a Reservoir Computing benchmark, although not as widespread as the dynamical systems tasks, which tend to translate better to the intrinsically temporal Reservoir Computing paradigm.

For non-stationary data tasks,
the high dimensional input may be provided in parallel,
as a single input vector, sequentially, as a non-stationary time series,
or a combination of both.
The categorical result is typically read out at a single time point after some time delay.

Tasks making a binary classification 
need only a single output node.
Tasks classifying into more than two categories
typically used multiple output nodes,
where each output node corresponds to one of the categories (see section~\ref{sec:class-cat-out}).

A style of classification task better suited to the medium of RC is time series based classification tasks, where the goal is to identify what the state of the time series is currently. This type of task has frequently been used in biomedical applications of RC, such as classifying gestures based on EMG signals \citep{Donati2018-km,Garg2021-jr}, identifying arrhythmic heartbeats \citep{Cucchi2021-cp}, or detecting seizures based on  EEG signals \citep{Kudithipudi2015-to}. No standard RC benchmark task for this style of classification appears to have yet emerged, however. 
We suggest that the Santa Fe sleep apnea data set would be a suitable candidate for an RC classification benchmark task (section~\ref{sec:Santa-Fe-apnea}).

\subsubsection{Property measures}

Rather than measuring performance on a particular task, some benchmarks instead measure more direct computational properties of the reservoir, such as its memory and generalisation capabilities. The object of these tasks is not to maximise a success rate or minimise an error rate. Instead, we can take these properties and see how the behaviour of the reservoir may make them better suited to different tasks.

\subsection{Other axes of classification}

We identify some other axes of benchmark classification:
these are not applicable in all cases.
\begin{itemize}
    \item known \textit{v} unknown dynamics
    \item supervised \textit{v} unsupervised learning
\end{itemize}

\subsubsection{Known \textit{v.} unknown dynamics}\label{sec:known}

Given the dynamical systems formalism,
many Reservoir Computing tasks involve training them to behave like other dynamical systems, producing the same time series outputs, in order to predict or to replace those others. 

Are the dynamics of the benchmark task known and understood by us, or are they unknown, with only the outputs and perhaps the inputs being observed? 
In the first case, we typically have a dynamical system model that we can use to generate the behaviour, such as NARMA (section~\ref{subsubsec:narma}) or Mackey--Glass (section \ref{sec:Mackey-Glass}). 
The second case typically uses observed natural phenomena, like laser data (section~\ref{sec:Santa-Fe}) or sunspot observations (section~\ref{sec:sunspots}), where we do not have a full model to generate the underlying dynamics, and are restricted to the supplied real world dataset.

\subsubsection{Supervised \textit{v.} unsupervised learning}

Most Reservoir Computers are trained through linear regression. This requires training data where the outputs are known, which is not always available. There are, therefore, some works in the literature that are focused on finding a way to train reservoirs that do not include the expected output. Two examples of unsupervised learning benchmarks 
 are channel equalisation (section~\ref{sec:unsup-chan-eq})
and pole balancing (section~\ref{sec:pole-bal}).

\section{Dynamics imitation tasks}\label{sec:imitate}

\subsection{NARMA: imitating known dynamics}\label{subsubsec:narma}

NARMA (Nonlinear Auto-Regressive Moving Average) is a family of imitation tasks widely used as benchmarks. It is based on two systems for modelling time series based dynamical systems~\citep{Weigend2018}, the Auto-Regressive (AR) and Moving Average (MA) models. 

Connor et al. \cite{Connor1991,Connor1994} introduce and define a generalisation of linear ARMA models; their generic NARMA$(p,q)$ model depends on $p$ past states and $q$ past inputs:
\begin{align}
    x(t) & = h(x(t-1), \ldots, x(t-p), 
    u(t-1), \ldots, u(t-q)) + u(t)
\end{align}
In their particular analysis, $h$ is an unknown smooth function to be estimated, 
and the input $u$ has zero mean.
They demonstrate a robust learning algorithm for approximating this function with a recurrent neural network (RNN).

Atiya and Parlos \cite[eqn.79]{Atiya2000-iq} introduce a second order NARMA system:
\begin{align}
    \label{eq:narma-2}
    x(t+1) &=  0.4 x(t) + {}
    0.4 x(t) x(t-1)  + 0.6 u^3(t) + 0.1
\end{align}
This is sometimes referred to in the literature as `NARMA2' \citep{Fujii2017-sx},
but is not of the same functional form as the family of equations usually referred to as NARMA-$N$.

Atiya and Parlos \cite[eqn.86]{Atiya2000-iq} also introduce a tenth order NARMA system,
which is now referred to as NARMA-10.
The functional form has been generalised \citep{Rodan2011-xb} as the $N$th order, or NARMA-$N$, system:
\begin{align}
    \label{eq:narma-equation}
    x(t+1) & = \alpha x(t) + {}
    \beta x(t) \left(\sum_{i=0}^{N-1}x(t-i) \right) +{}
    \gamma u(t - N+1)u(t) + \delta
\end{align}
Its non-linearity comes from the various $x(t)x(t-i)$ terms, 
and the $u(t - N+1)u(t)$ term;
it also requires memory of the previous $N$ states.


NARMA-10 using the original parameter values and the input stream generated from a uniform random distribution in the range [0, 0.5] is one of the commonest usages in the literature; see for example
\cite{Verstraeten2007-wc,Rodan2010-ku,Holzmann2010-mo,Paquot2012-li,Goudarzi2015-gv,Vinckier2015-td,Inubushi2017-zn,Dale2018-se}.

\begin{figure*}
    \centering
    \includegraphics[trim = 9mm 4mm 12mm 14mm, clip, width=\linewidth]{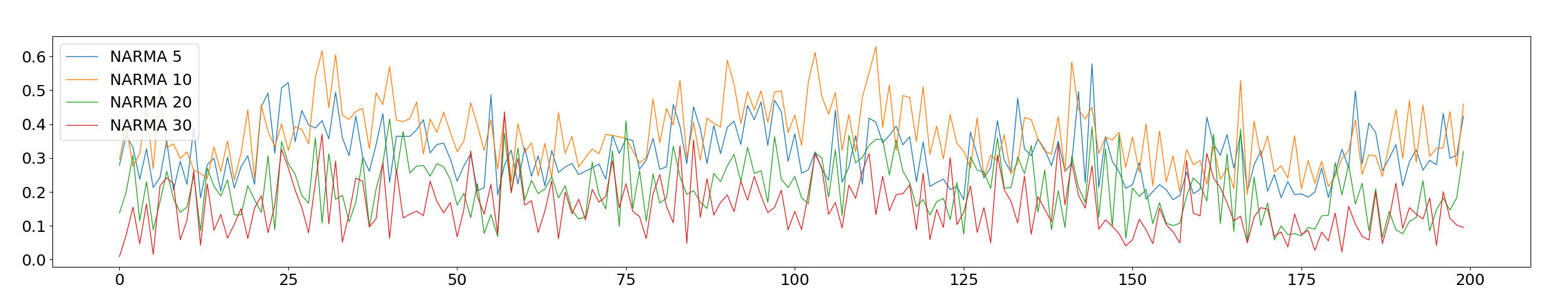}
    \caption{
    200 timesteps of NARMA-5, NARMA-10, NARMA-20 (with tanh), and NARMA-30, 
    using the starred parameter values from table~\ref{table:narma-equations}, 
    and the same input stream $u(t)$ drawn from $U[0,0.5]$.
    }
    \label{fig:narmas}
\end{figure*}

\begin{table*}[tp]\small
\begin{center}
\begin{tabular}{rllllccl}
\toprule%
$N$ & $\alpha$ & $\beta$ & $\gamma$ & $\delta$ & input & fig.\ref{fig:narmas} & introduced in  \\
\midrule
{5} & {0.3} & {0.05} & {1.5} & {0.1} & $U[0,0.2]$ && {\cite[eqn.18]{Fujii2017-sx}} \\
{5}  & {0.3} & {0.05}  & {1.5}  & {0.1} & $U[0,0.5]$&*& {\cite[chap.3]{Dale2018-rr}} \\[2mm]
{10} & {0.3} & {0.05} & {1.5} & {0.1} & $U[0,0.5]$ &*& {\cite[eqn.86]{Atiya2000-iq}} \\[2mm]
{15} & {0.3} & {0.05} & {1.5} & {0.1} & $U[0,0.2]$&& {\cite[eqn.18]{Fujii2017-sx}} \\[2mm]
{20} & {0.3} & {0.05} & {1.5} & {0.01} & $U[0,0.5]$&*& {\cite[eqn.6]{Rodan2011-xb}, with tanh(.)} \\
{20} & {0.3} & {0.05} & {1.5} & {0.1} & $U[0,0.2]$&& {\cite[eqn.18]{Fujii2017-sx}} \\[2mm]
{30} & {0.2} & {0.04} & {1.5} & {0.001} & $U[0,0.5]$ &*& {\cite[sec.3]{noauthor_2008-fe}} \\
{30} & {0.2} & {0.004} & {1.5} & {0.001} & $U[0,0.5]$&& {\cite{Dale2018-se}} \\
\bottomrule
\end{tabular}
\end{center}
\caption{
Summary of NARMA-$N$ parameters found in literature. 
The NARMA sequence is plotted for the starred values in Figure~\ref{fig:narmas}
}
\label{table:narma-equations}
\end{table*}
Adapting this for different values of $N$ reveals that the equation can rapidly diverge if not tuned correctly.
Indeed, even the standard NARMA-10 setup does itself occasionally diverge
\citep{Kubota2021}.
Other authors have defined the form for further values of $N$, and taken a variety of approaches to controlling divergence.
For example, 
Schrauwen et al. \cite[p.1164]{noauthor_2008-fe} define NARMA-30 with different parameter values of $(0.2, 0.04, 1.5, 0.001)$, nevertheless, it too very occasionally diverges);
Rodan and Tino \cite[eqn.6]{Rodan2011-xb} define NARMA-20 and add a $\tanh(.)$ wrapper to stop divergence;
Fujii and Nakajima \cite[eqn.18]{Fujii2017-sx} define NARMA-5, 10, 15, and 20 
with the original parameter values but a restricted input range  $u \in\mbox{\textsc{Uniform}}[0,0.2]$.
Parameter values for these and other cases from the literature are given in table~\ref{table:narma-equations}.

Furthermore, various NARMA time series each have their own range of output $x$ values.
For example, 
Atiya and Parlos \cite{Atiya2000-iq}'s NARMA-10 $x$ values range between (ignoring divergences) 0.15 and 1,
Fujii and Nakajima \cite{Fujii2017-sx}'s NARMA-10 with reduced $u$ range has $x$ values ranging between 0.15 and 0.25,
and d Schrauwen et al. \cite{noauthor_2008-fe}'s NARMA-30 $x$ values range between (again ignoring divergences) 0 and 0.6.
Even using consistent parameter values and input range yields systematically different output ranges for different values of $N$.
This makes any performance comparison as a function of $N$ problematic.

The relative simplicity of the task, and the ability to parameterise the difficulty by using a range of $N$ values, has made this benchmark popular when evaluating small reservoirs, as is often true for \textit{in materio} reservoir computers~\citep{Dale2018-rr}.
As noted, however, there are several issues: 
the potential for divergence when generating the series, a multitude of parameter values in the literature making comparisons difficult,
and no clear trends as a function of $N$.
As such, its use as a benchmark should be approached with caution.
NARMA is a non-linear memory task,
so using the generic, more readily interpretable, albeit more computationally intensive, Information Processing Capacity (IPC) measure (section \ref{sec:IPC}) can overcome these limitations.

Authors using this benchmark typically use it alongside other benchmarks, such as the Santa Fe Laser Task~\citep{Inubushi2017-zn} (sec.\ref{sec:Santa-Fe}), the Mackey--Glass System~\citep{Goudarzi2015-gv} (sec.\ref{sec:Mackey-Glass}) or the Multiple Superimposed Sines task~\citep{Holzmann2010-mo} (sec.\ref{sec:mso}).

\subsection{Channel equalisation: imitating unknown dynamics}
\label{subsubsec:channel-equalisation}

The Channel Equalisation task is one that involves restoring a noisy signal to its original state. It is a task that is chosen for its real-world applicability.

Channel Equalisation can be modelled as an imitation task: the Reservoir Computer imitates a `perfect' filter, which completely removes the noise. This filter is unknown and may even be impossible, but as the output is known (the noiseless signal) and the training relies solely on the output, it can still be imitated to some degree. 
This task may be evaluated using the symbol error rate~\citep{Jaeger2004-eu}.

Jaeger and Haas \cite{Jaeger2004-eu} first applied the task to Reservoir Computing. A description of the method used can be found in their supplementary materials \citep{Jaeger2004-vb}. 
The channel model used in \cite{Jaeger2004-eu} was first introduced by Mathews and Lee \cite{Mathews1994-df}. 

\begin{figure}
\centering
    \includegraphics[trim = 27mm 6mm 26mm 12mm, clip, width=0.7\linewidth]{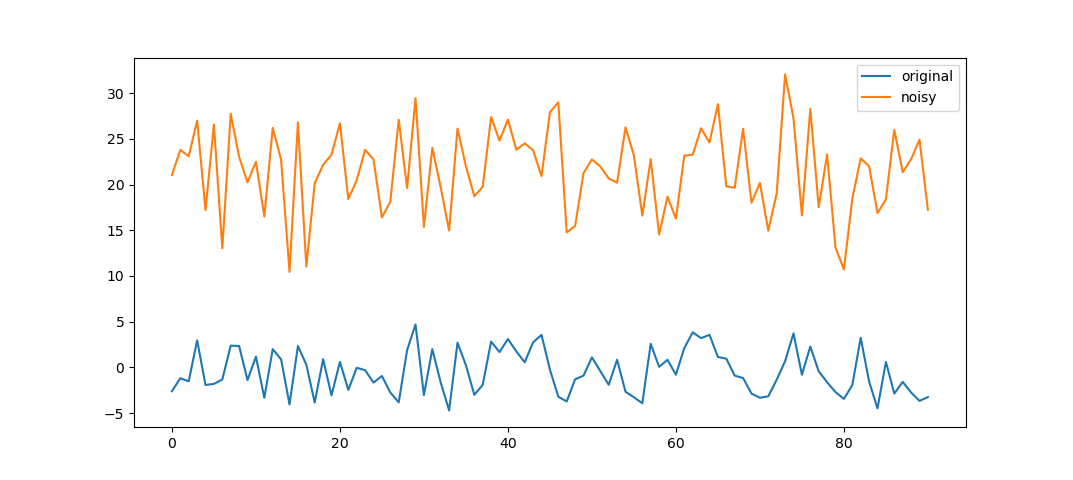}
    \caption{
    A channel signal (bottom, blue), and the signal with nonlinear noise applied (top, orange).}
    \label{fig:noisy-channel}
\end{figure}

The system takes as input a signal to which nonlinear noise has been added:
the task is to output the signal with the noise removed. 
Jaeger \cite{Jaeger2004-vb}
uses a signal $d(n)$ that is a sequence of randomly chosen values from the set $\{-3, -1, 1, 3\}$. 
The signal $d(n)$ is passed through a linear channel\footnote{In \cite{Jaeger2004-vb} the coefficient is written as 0.09 1; however, both the original~\citep{Mathews1994-df} and the `other formulation'~\citep{Rodan2010-ku} use 0.09, indicating that \cite{Jaeger2004-vb} may have a typographical error.}:       
\begin{align}
    q(n) &= 0.08 d(n + 2) - 0.12 d(n + 1) +  {}
    d(n) + 0.18 d(n - 1) - 
    0.1 d(n - 2)  
        \\      \nonumber
    & {}+     
    0.09 d(n - 3) - 0.05 d(n-4) + 
    0.04 d(n - 5) + {} 
    0.03 d(n - 6) +
    0.01 d(n-7)
    \label{eq:chan-lin}
\end{align}
Then nonlinear noise applied:
\begin{equation}
    u(n) = q(n) + 0.036 q(n)^2 - 0.011 q(n)^3 + v(n)\label{eq:chan-nlin}
\end{equation}
where
$v(n)$ is Gaussian white noise, applied to ensure that the signal-to-noise ratio of the output is between 12 and 32 decibels~\citep{Jaeger2004-eu}.\footnote{%
\cite{Jaeger2004-eu} specifies 12 to 32 dB;
the supplementary material~\citep{Jaeger2004-vb} specifies 16 to 32 dB. 
Most authors who use the benchmark~\citep{Vinckier2015-td, Paquot2012-li, Antoine_Dejonkheere_Francois_Duport_Anteo_Smerieri_Li_Fang_Jean-Louis_Oudar_Marc_Haelterman_Serge_Massar2014-dm, Duport2012-cw}
use the lower bound of 12 dB.
There are some other formulations;
for example~\cite{Rodan2010-ku} has no noise term and adds an offset of  +30.
} An example of such a signal, as well as the signal with noisy applied, can be found in figure \ref{fig:noisy-channel}.

One can also find certain tasks in the literature that are called `Channel Equalisation', but do not follow the same methodology as the best known version of the benchmark. 
For example, Boccato et al. \cite{Boccato2011-ke} use a similar task, but the signal is composed solely of the values {–1, +1}, and the evaluation is done via Average Mean Squared Error. 
Boccato et al. \cite{Boccato2012-pc} follow this up by adding an unsupervised channel equalisation task, where the noiseless signal is not known. While this task has a similar purpose to the classic, supervised version of the task, the lack of “clean” signal used in training signifies that it is not an imitation task. This provides us with an interesting distinction between the form of the task (imitation of an unknown dynamical system, or not) and its function (channel equalisation, in both cases).

\subsubsection{Unsupervised Channel Equalisation}\label{sec:unsup-chan-eq}
An unsupervised version of the Channel Equalisation task exists \citep{Boccato2012-pc}. The argument is that most distorted signals do not have a “clean” version that can be used to train the nonlinear filter, therefore it is useful to have a version of the task that uses unsupervised learning as opposed to imitation-based learning.
Instead of training the received signal against the perfect, noiseless signal, the reservoir was trained against the information conveyed. The authors argue that, since the explicit signal itself was not used in training, this qualifies the task as an unsupervised learning task.
As the `perfect signal' is still known, however, the task is evaluated against that perfect signal.

\subsection{Two-jointed arm: imitating unknown dynamics}\label{sec:2jointarm}

The two-jointed arm task~\citep{Slotline1991-qw} involves controlling a two-jointed (robotic) arm in order to move it from point A to point B. 
Like the Channel Equalisation task, this task can be seen as an imitation of an unknown, perfect controller, of which only the outputs are known. 

Joshi and Maass \cite{Joshi2004-pc,Joshi2005-md} use this task in the context of Reservoir Computing.
The training data used are points along the trajectory that the robotic arm was intended to follow. The robotic arm is then tested in a `closed loop', where no training data is available, what we in section~\ref{sec:imitate-intro} refer to as `in-situ' testing.  

The arm moves from point A to point B along a horizontal plane, allowing gravitational forces to be ignored. The dynamics of the arm are modelled by:
\begin{equation}
    \label{eq:arm-dynamics}
    \mathbf{H}(\boldsymbol\theta)\ddot{\boldsymbol\theta} 
    + \mathbf{C}(\boldsymbol\theta, \dot{\boldsymbol\theta})\dot{\boldsymbol\theta} 
    = \boldsymbol\tau
\end{equation}
where $\boldsymbol\theta = [\theta_1, \theta_2]^T$ are the angles of the arm's joints, $\boldsymbol\tau = [\tau_1, \tau_2]^T$ are the joint input torques,
$\mathbf{H}$ is a $2\times2$ inertia matrix, and $\mathbf{C}$ is a $2\times2$ matrix of the Coriolis and centripetal terms. The reservoir is then given a target trajectory, for which it must imitate a `perfect' controller in a model-free manner. 
See \cite{Joshi2004-pc,Joshi2005-md} for details.


In this context, we see the advantage of testing the task \textit{in situ}: the results show that, despite not following the same trajectory presented in the training data, the robotic arm still reaches its destination.

\subsection{Van der Pol oscillator: imitating unknown dynamics of a controller}
\label{sec:van-der-pol}

The Van der Pol (VdP) oscillator was first described by the Dutch physicist Balthazar Van der Pol in the 1920s~\citep{van1922lxxxv} and later employed to model the oscillations of the heart~\citep{van1928lxxii}. 
It is particularly useful for modelling systems that exhibit non-linear behaviour, such as relaxation oscillations and self-sustained oscillations. 
Within the context of machine learning, the VdP oscillator is commonly used to test the effectiveness of neural networks~\citep{nicola2017supervised,yeo2019data}.  
The VdP oscillator can be applied as the substrate of physical reservoir computing \cite{shougat2023van}.

\begin{figure}[tp]
\centering
\subfloat[$x_0 = -1$, $\dot{x_0}$ = 1.5\label{subfig-1:NoPID x-1 xd 1.5}]{%
  \includegraphics[clip, trim=6.5cm 8.5cm 6.5cm 9cm,width=0.3\textwidth]{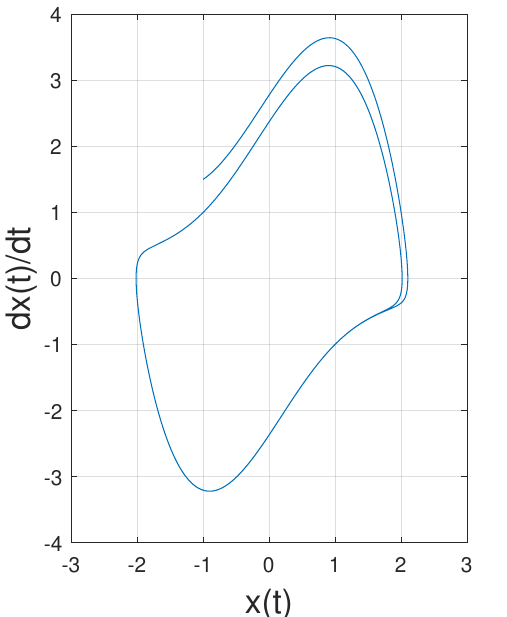}}
  \quad
\subfloat[$x_0$ = 0.5, $\dot{x_0}$ = 0.75\label{subfig-2:No PID x0.5 xd 0.75}]{%
  \includegraphics[clip, trim=6.5cm 8.5cm 6.5cm 9cm,width=0.3\textwidth]{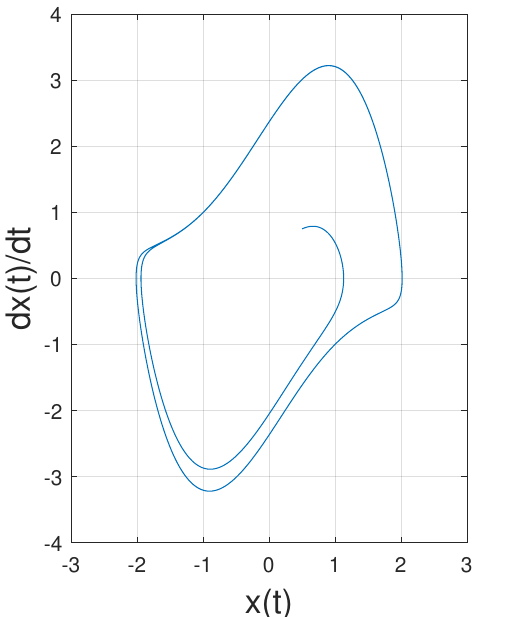}}
\caption{Phase plane of VdP oscillator in Equation~\ref{Forced VdP equation} with \textit{F(t)} = 0 (no external force applied). Parameter settings: $\mu$ = 1.5, the initial condition of each plot is given.}
\label{fig:phase plane of VdP}
\end{figure}

The VdP oscillator is characterised by a nonlinear damping term, and the forced VdP with an external force $F(t)$, is expressed as a nonlinear differential equation:
\begin{equation}
\label{Forced VdP equation}
    \frac{d^{2}x}{dt^{2}}-\mu(1-x^2)\frac{dx}{dt}+x=F(t)
\end{equation}
where $x$ is the state variable, $t$ is time, and $\mu$ is a scalar parameter that dictates the nonlinearity and the strength of the damping.
$F(t)$ is the external force applied to the VdP system. In the undriven VdP system the external force term is zero $F(t)=0$, with the phase plane trajectory as shown in Figure~\ref{fig:phase plane of VdP}.

The VDP task can be set up as an imitation task, where a reservoir is trained to imitate a PID controller. Once trained, the reservoir controller is tested in situ, where the output is compared to the target trajectory as opposed to the output of the PID controller. 

In a recent study, the VdP oscillator is used as a model-free control task where a reservoir is trained to produce an external force to make the VdP system produce a circle \citep{10191630,gan2024}.


\subsection{Pole balancing : unsupervised learning}\label{sec:pole-bal}
Pole balancing is a family of benchmark tasks for controllers. It involves simulating a cart moving back and forth along a single axis, trying to keep one or several poles that are attached to the cart by a hinge upright. The task is considered a success if the poles remain upright for a certain amount of time.

In the field of Reservoir Computing, this task is primarily used to evaluate unsupervised learning in reservoirs.
It has been used for exploring different methods of training reservoirs, such as neuroevolution \citep{Jiang2008-rm, Jiang2008-im} and NEAT \citep{Chatzidimitriou_undated-hy}.
One could also implement this task as an imitation task, and have the reservoir imitate the unknown dynamics of a perfect controller. 

While there are multiple formulations of the pole balancing task, double pole balancing without velocity information is the most popular in this context. 
Given a cart that can move along the $x$ axis, on which are mounted two poles by a hinge, the goal is to move the cart such that the poles stay balanced upright. 
The poles are of different lengths and masses.\footnote{An implementation of the pole balancing problem that is frequently used in the literature is that of Kenneth O. Stanley, and can be found at \url{http://nn.cs.utexas.edu/?dpb-esp}}
At each timestep, the Reservoir Computer outputs a force 
to be applied to the cart. 
The task is considered a success if the poles stay upright for 100,000 steps, which is equivalent to roughly thirty minutes of simulation time.

\section{Dynamics prediction tasks}\label{sec:predict}

In this section, we describe the systems commonly used as prediction benchmarks in Reservoir Computing. A greater variety of datasets that can be used for benchmark tasks can be found in Gilpin \cite{DBLP:journals/corr/abs-2110-05266}'s database of chaotic dynamical systems, all specified according to guidelines proposed by Gebru et al. \cite{gebru2021datasheets}.

\subsection{Mackey--Glass equation : known dynamics}\label{sec:Mackey-Glass}

The Mackey--Glass benchmark is derived from a delay differential equation  introduced to model certain physiological systems \citep{Micheal_C_Mackey1977-vr}.\footnote{%
This delay differential equation is also used elsewhere in Reservoir Computing to create a nonlinear node in delay line reservoirs (see, for example, \cite{Appeltant2011-el}).  This other usage is not examined here.
}
These models characterise illness through changes to the variables that lead to these systems turning chaotic.
The authors examine various models, one for the breathing patterns of patients with Cheyne-Stokes illnesses, and two variants for the growth rates of white blood cells in leukaemia patients.  

The latter model is used as the basis for the benchmark; 
\cite[eqn.4b]{Micheal_C_Mackey1977-vr} states:
\begin{equation}
    \dot{P} = \frac{\beta \theta^n P_{t-\tau}}{\theta^n + P^n_{t-\tau}} - \gamma P_t
\end{equation}
where $P_t >0$ is the concentration of circulating blood cells at time $t$;
$\tau$ is the time delay between initiating blood cell production and the mature blood cells being released;
$\beta > 0$ is the base level production rate of cells; 
$\theta > 0$ is the baseline concentration; 
$n >0$ is a real-valued non-linearity parameter;
$\gamma >0$ is the decay rate of cells.

\begin{figure}[tp]
    \begin{center}
    \includegraphics[width=0.7\linewidth]{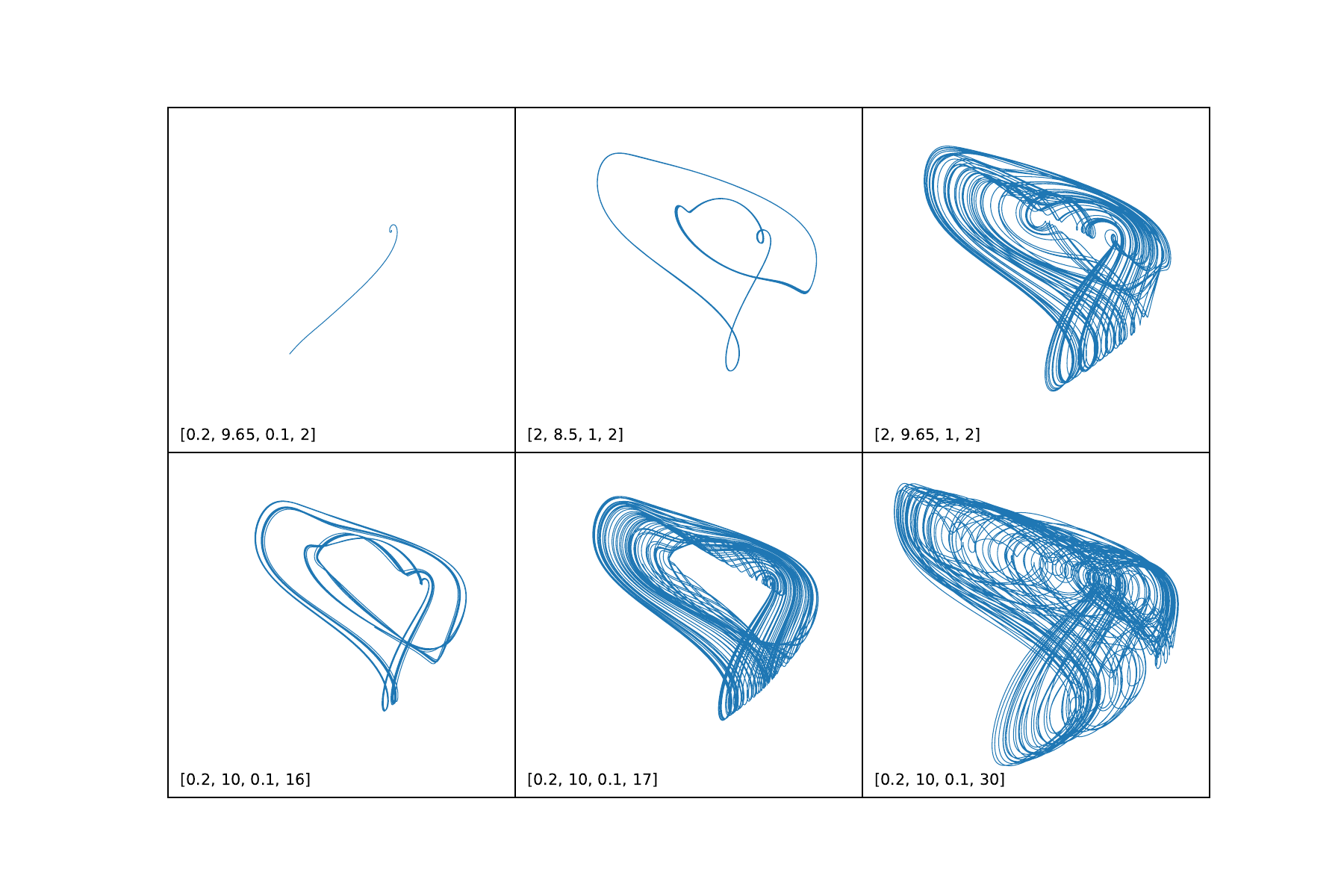}
    \caption[Time delay embedding plots of the Mackey--Glass equation]{\small
    Time delay embedding plots of the Mackey--Glass equation, for various parameter values.  
    Each figure is a plot of $x(t-\tau)$ against $x(t)$ for a particular set of parameter values 
    $[\beta, n, \gamma, \tau]$.
    The $x$ and $y$ axes of all plots run from 0.1 to 1.5.
    Plotted for $150\tau$ time units after the transient,
    and an initial condition $x(t<0) = 0.5$.
    The transient of the first 100 time units are not plotted (except for the top left plot, where the entire transient is shown).
    \\Top: $[\beta, n, \gamma, \tau]$ values from \cite{Glass2010}:
    \\
    $\bullet~[0.2, 9.65, 0.1, 2]$ : transient to a point attractor
    \\
    $\bullet~[2, 8.5, 1, 2]$ :  periodic behaviour [their fig.5]
    \\
    $\bullet~[2, 9.65, 1, 2]$ :  chaotic behaviour [their fig.2,fig.7]
    \\Bottom: $[\beta, n, \gamma, \tau]$ values from \cite{Jaeger_undated-ek}:
    \\
    $\bullet~[0.2, 10, 0.1, 16]$ :  periodic behaviour 
    \\
    $\bullet~[0.2, 10, 0.1, 17]$ :  mildly chaotic behaviour
    \\
    $\bullet~[0.2, 10, 0.1, 30]$ :  wildly chaotic behaviour
    \\
    (Plots produced using the python {\sf ddeint}
    delay differential equation solver, {\url{https://pypi.org/project/ddeint/}}
    with an integration timestep of 0.02 time units.)
    }\label{fig:mackey-glass}    
    \end{center}
\end{figure}

Normalising the concentration with respect to the baseline, $x_t = P_t/\theta$, gives the more usual form of the  Mackey--Glass chaotic equation \citep[eqn.1]{Glass2010}:
\begin{equation}\label{eqn:mackey-glass}
    \dot{x} = \frac{\beta x_{t-\tau}}{1 + x^n_{t-\tau}} - \gamma x_t
\end{equation}
Here $x$ is a dimensionless normalised quantity, expressed in `units' of the parameter $\theta$.
$\theta$ is still a parameter of the model, but is now implicit in the equation: it provides the scale for the dependent variable $x$.
Eqn.\ref{eqn:mackey-glass} has chaotic dynamics for 
$\beta=2$, $n=9.65$, $\gamma=1$, $\tau=2$
with an initial condition of $x=0.5$ for $t<0$ \citep{Glass2010}.\footnote{%
The text of \citep{Glass2010} states $\beta=0.2, \gamma=0.1, \tau = 2$, 
but the captions of its figures 1 and 2 state $\beta=2, \gamma=1, \tau = 2$.
Figure~\ref{fig:mackey-glass} here favours the values from the captions.
Equivalently, one could use $\beta=0.2, \gamma=0.1, \tau = 20$.
}

Jaeger \cite{Jaeger_undated-ek} states that the Mackey--Glass system is often used for learning dynamical systems from data, `invariably' with parameter values of $\beta=0.2, n=10, \gamma=0.1$, and that with
these values the system is chaotic for $\tau > 16.8$;
see \citep{Farmer1982} for a detailed analysis.
Jaeger \cite{Jaeger_undated-ek} investigates $\tau=17$ (mildly chaotic) and $\tau=30$ (which he dubs `wildly' chaotic).
See figure~\ref{fig:mackey-glass}.

The parameter values used in \cite{Glass2010} and  \cite{Jaeger_undated-ek} appear to differ considerably (figure~\ref{fig:mackey-glass}); however, timescale rescaling shows they are comparable.
There are three timescales in eqn.\ref{eqn:mackey-glass}:
$1/\beta$ (production, or growth, timescale);
$1/\gamma$ (decay timescale),
and $\tau$ (time delay).
If we normalise these with respect to the decay timescale,
we have two independent timescales: $\gamma/\beta$ and $\gamma \tau$.
\cite{Glass2010} uses $\gamma/\beta = 1/2$, $\gamma\tau = 2$;
\cite{Jaeger_undated-ek} uses $\gamma/\beta = 1/2$,  $\gamma\tau= 1.7, 3.0$.
\cite{Glass2010} explores the transition between periodic and chaotic behaviour
using fixed timescales and varying the non-linearity parameter $n$
(the timescales have biological meaning,
whereas the non-linearity is a parameter of the model).
\cite{Jaeger_undated-ek} fixes the non-linearity at $n=10$ and instead varies the delay feedback timescale $\tau$; here the model is being used in a more abstract manner, and varying the time delay is a standard approach to investigating chaotic behaviours in DDEs \citep{Otto2019}.

Eqn.\ref{eqn:mackey-glass} describes a continuous system.
Jaeger \cite{Jaeger_undated-ek} describes a discretisation process for converting it to a discrete system suitable for generating time series data.
\begin{itemize}
    \item 
Discretise eqn.\ref{eqn:mackey-glass} using $dx/dt \approx (x(t+\Delta t) - x(t)) / \Delta t$:
\begin{equation}\label{eqn:mackey-glass-xhat}
    \hat{x}(t+\Delta t) = \hat{x}(t) + \Delta t \left( \frac{\beta \hat{x}(t-\tau)}{1 + \hat{x}^n(t-\tau)} - \gamma \hat{x}(t) \right)
\end{equation}
    \item Take $1/ \Delta t = N$, an integer (\cite{Jaeger_undated-ek} uses $N=10$),
to produce the time series
$
    \hat{x}(t), \hat{x}(t+\Delta t), \hat{x}(t+2\Delta t), \ldots \hat{x}(t+(N-1)\Delta t), \hat{x}(t+1), \ldots
$

    \item Subsample the full series every $N$ steps to produce the reduced series
$
    \hat{x}(t), \hat{x}(t+1), \hat{x}(t+2), \ldots
$

    \item Finally, transform the individual values to the interval $[-1,1]$:
\begin{equation}\label{eqn:mackey-glass-z}
    y(t) = \tanh(\hat{x}(t) - 1)
\end{equation}
\end{itemize}
The time series $y(t), y(t+1), \ldots$ provides the training data for the benchmark.

Using this discretisation method, \cite{Jaeger_undated-ek} generates two  sequences for each value of $\tau =17$ and $30$,
one sequence of length 3,000 and of 21,000 (four sequences in total).
These form the combined washout and training sequences.
The ESN
is trained on this data,
then tested on its ability to predict the 84th output value, compared to the expected 84th value, computing the relevant Normalised Mean Squared Error (NMSRE$_{84}$) of the value over multiple runs. 

Mackey--Glass is used as a benchmark by Jaeger again in \citep{Jaeger2004-eu}, this time with a single 3,000-step training sequence, and the NMSRE of the 84th subsequent value taken over 100 runs. The $\tau$ values used are not specified.
Holzmann and Hauser \cite{Holzmann2010-mo} also use the system as a benchmark, using a method almost identical to \citep{Jaeger_undated-ek}, though with the NMSRE taken over 100 runs, and an additional NMSRE of the 120th value used for the training sequence of length 21,000.
Roeschies and Igel \cite{Roeschies2010-rj} base their method on \citep{Jaeger_undated-ek}, this time using a single sequence of length 3,000 with $\tau=17$. One other notable difference is the use of a washout of 100 values, whereas prior works have used washouts of 1,000.
\cite{Goudarzi2015-gv} and \cite{Moon2019-du} depart from this traditional setup in order to propose their own.
These various different uses are detailed in table \ref{table:mackey-glass-lengths}.

\begin{table*}[tp]\small
\centering
\begin{tabular}{lcrcrrr}
\toprule%
source & \specialcell{no. of\\test seqs} & \specialcell{seq\\length} & $\tau$ value & \specialcell{washout\\length} & test value& \specialcell{no. of\\runs}\\
\midrule
\cite{Jaeger_undated-ek}& 1 & \specialcell{3000} & 30 & 1000 & 84 & 50\\
--''--& 1 & \specialcell{21000} & 30 & 1000 & 84 & 50\\
--''--& 1 & \specialcell{3000} & 17 & 1000 & 84 & 20\\
--''--& 1 & \specialcell{21000} & 17 & 1000 & 84 & 20\\
\cite{Jaeger2004-eu} & 1 & 3000 & ? & 1000 & 84 & 100\\
\cite{Holzmann2010-mo} & 2 & \specialcell{3000} & 17, 30 & 1000 & \specialcell{84} & 100\\
--''--& 2 & \specialcell{21000} & 17, 30 & 1000 & \specialcell{120} & 100\\
\cite{Roeschies2010-rj} & 1 & 3000 & 17 & 100 & 84 & 50\\
\cite{Goudarzi2015-gv} & 1 & 2000 & ? & ? & 2000 & 10\\
\cite{Moon2019-du} & 1 & 500 & 18 & ? & 50--52 & ?\\
\bottomrule
\end{tabular}
\caption{
A summary of different parameters used in the literature for the Mackey--Glass prediction benchmark. Values marked with `?' are not reported in the cited work. 
\cite{Goudarzi2015-gv} uses NMSE rather than NMSRE.
}\label{table:mackey-glass-lengths}
\end{table*}

This benchmark is an ideal one to use in prediction of known systems, not only because its past use is prevalent and well documented, but also because, as a chaotic system, it is a rather challenging task, particularly when certain parameter values are set, such as $\tau \geq 16.8$. The use of the NMSRE over a given number of runs on a single value is unusual, however, and should be highlighted to avoid confusion.

\subsection{Multiple Superimposed Oscillators: known dynamics}\label{sec:mso}
The Multiple Superimposed Oscillator task, sometimes also referred to as the Multiple Superimposed Sines, or the Multiple Sines task, is a known dynamical system prediction benchmark.

The object of the task is to imitate a sequence generated by a sum of sines with incommensurate frequencies, and hence with a very long overall period, particularly for large $n$:
\begin{equation}
   x(t) = \sum_{i=1}^{n} \sin(\alpha_i t)\label{eq:mso}
\end{equation}
where the frequencies $\alpha \in [0.2,$ $0.311,$ $0.42,$ $0.51,$ $0.63,$ $0.74,$ $0.85,$ $0.97]$. 

The task was originally introduced in a presentation by Jaeger \cite{Jaeger2004},
for $n=2$  (what would now be called MSO-2, but there called `additive dynamics'), as an example of a task that an ESN finds `impossible to learn'.
The task difficulty has been increased
by extending the list of $\alpha$ with higher frequency values to give MSO-5 \cite{Wierstra2005}
and MSO-8 \cite{Roeschies2010}.
Other uses of this benchmark include \cite{Xue2007-df} (MSO-2)
and \cite{Koryakin2012-og} (MSO-8).

Since Jaeger \cite{Jaeger2004} states that
the task cannot be solved by the standard ESN model,  
some authors use this benchmark on extensions to the ESN model \citep{Wierstra2005,Schmidhuber2007-ux,Xue2007-df,Koryakin2012-og}, for example, by evolving the weights, or using feedback, to demonstrate superior performance.
However, if one uses just the standard ESN model with similar training and testing dataset lengths as these variants (many hundreds of timesteps each), one can also achieve apparently reasonable NRMSE values (see section~\ref{sec:NRMSE}),
implying that the ESN has learned the data.
Jaeger et al. \cite[fig.4]{Jaeger2007-mb} demonstrate that this good prediction does not hold indefinitely:
 after many \textit{thousands} of timesteps, the prediction diverges.
This is what Jaeger \cite{Jaeger2004} means when referring to the task as `impossible to learn'.

\subsection{Lazy Figure 8: known dynamics}\label{sec:lazy8}
The figure 8 task is a `perennial task' in the field of Recurrent Neural Networks \citep{Jaeger2007-mb}, and a challenging one.
It involves predicting the next point along a sequence that traces the shape of the digit 8, and as such is a prediction of a known system.

In Reservoir Computing, this task is adapted to the `Lazy figure 8'\footnote{%
While the `Lazy' in `Lazy Eight' typically refers to the orientation of the figure 8 drawn on its side ($\infty$), Jaeger instead calls it thus due to the timescale over which the figure is drawn.
} \citep{Jaeger2007-mb}, where the full shape is composed of 200 points. Training is performed over 3000 steps, which includes a washout period of 1000 steps. Two experiments were performed, the first where the 200 points were equally spaced over the figure, and a second `time-warped' where the spacing of the points varied over time.

Like the MSO task, this benchmark is one that cannot be satisfactorily performed by a classical ESN, although Jaeger et al. \cite{Jaeger2007-mb} show that their leaky integrator based model is capable of accomplishing it. 
This is an example of a benchmark being used to show that a newer model performs better than classical ESNs.

\begin{figure}[tp]
    \begin{center}
    \includegraphics{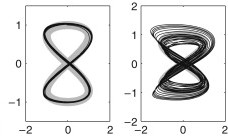}
    \caption{
    Lazy Figure 8 \citep[fig.6]{Jaeger2007-mb}. This figure follows a trajectory of 200 points, which is then repeated over a training and subsequent testing period. The two figures are of the output from the ESNs, with (left) constant and (right) `time-warped' inputs. 
}
    \label{fig:lazy-eight}
    \end{center}
\end{figure}

\subsection{Lorenz chaos: known dynamics}\label{sec:lorenz}
Lorenz chaos refers to two systems of equations initially derived to model the behaviour of certain weather-based dynamical systems. The task is to predict the behaviour of the equational model, and the benchmark is thus a prediction of a known system.
These systems have chaotic behaviour, where small differences in state can lead to large divergences in behaviours. This makes predicting it an interesting task in Reservoir Computing.

\subsubsection{Lorenz'63 system}
\begin{figure}[tp]
\begin{subfigure}{0.5\textwidth}
\centering
    \captionsetup{width=.9\linewidth}
    \includegraphics[scale=0.45,clip,trim=3.5cm 1cm 2cm 1.7cm]{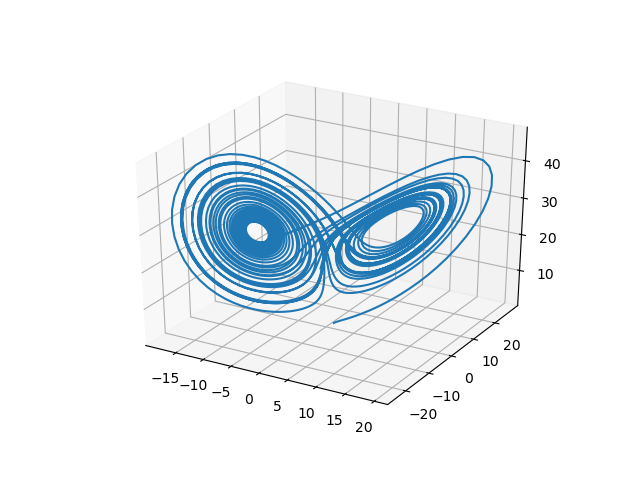}
    \caption{
    Lorenz'63 model (eqn.\ref{eq:lorenz-63}) with  $\sigma=10$, $\rho=28$, $\beta=8/3$.}
    \label{fig:lorenz-63}
\end{subfigure}
\begin{subfigure}{0.5\textwidth}
\centering
    \captionsetup{width=.9\linewidth}
    \includegraphics[scale=0.45,clip,trim=3.5cm 1cm 2cm 1.7cm]{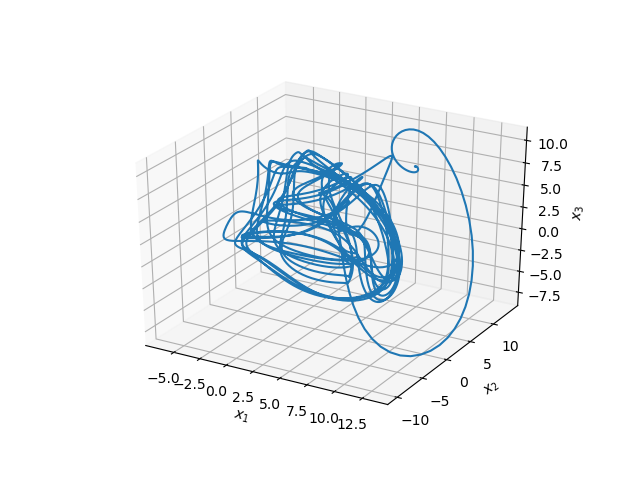}
    \caption{
    Projection of a Lorenz'96 model (eqn.\ref{eq:lorenz-96}) with  $N=5$, $F=8$}
    \label{fig:lorenz-96}
\end{subfigure}
\caption{
Illustrations of the two Lorenz systems.}\label{fig:lorenz}
\end{figure}

The first system, introduced in 1963 \citep[eqn.25--27]{DeterministicNonperiodicFlow}, is frequently referred to as Lorenz'63.
The set of three coupled ODEs are:
\begin{equation}
    \label{eq:lorenz-63}
    \begin{split}
    \dot{x} &= \sigma(x - y)\\
    \dot{y} &= x(\rho - z) - y\\
    \dot{z} &= xy - \beta z\\
    \end{split}
\end{equation}
The values of the parameters $\sigma$, $\rho$, and $\beta$
change behaviour between periodic and chaotic.
The parameter values used are typically $\sigma = 10$, $\rho = 28$,  $\beta = 8/3$ \citep{Pathak2017-fj,Roeschies2010-rj,Canaday2021-bh}.
The behaviour for this chaotic choice is illustrated in figure \ref{fig:lorenz-63}.

\subsubsection{Lorenz'96 system}
The second system, introduced in 1996 \citep[eqn.1]{Lorenz_undated-wb}, is typically referred to as Lorenz'96. 
It is described by $N$ variables $x_i$, $i\in 0\ldots N-1$, governed by $N$ coupled ODEs:
\begin{equation}
    \dot{x}_i = (x_{i+1} - x_{i-2})x_{i-1} - x_i + F
    \label{eq:lorenz-96}
\end{equation} 
where the index arithmetic is modulo $N$,
and the constant $F$ is independent of $i$.
For small $F$, the solution is $x_i = F$ for all $i$;
for intermediate values of $F$ there are periodic solutions;
for larger values of $F$ solutions are chaotic.
(The value of $F$ that marks the transition to chaos depends on $N$.)
The behaviour for one choice of $N$ and $F$ is shown in  figure \ref{fig:lorenz-96}.
Table \ref{table:lorenz-96-params} gives values of $N$ and $F$ used in the literature.

\begin{table}[tp]
\begin{center}
\begin{minipage}{174pt}%
\begin{tabular}{@{}lcc@{}}
\toprule
 source & $N$ & $F$\\
\midrule
  \cite{Vlachas2018-ol} & 0--40 & 4--16\\
  \cite{Vlachas2019-rq} & 0--40 & 8, 10\\
  \cite{Canaday2021-bh} & 11 & 8\\
\bottomrule
\end{tabular}
\end{minipage}
\caption{
Parameter values used when generating a Lorenz'96 sequence.}\label{table:lorenz-96-params}
\end{center}
\end{table}

\subsection{Kuramoto--Sivashinski equation: known dynamics}\label{sec:ks-eqn}

The Kuramoto--Sivashinsky  (KS) equation is a fourth order PDE originally derived to model diffusion-induced chemical turbulence \citep{Sivashinsky1977-tl, Sivashinsky1980-qd}. When used as a benchmark, the task is to predict the next output of the system, and therefore is a prediction task for a known system.

\begin{figure}
    \centering
    \includegraphics[width=0.7\linewidth]{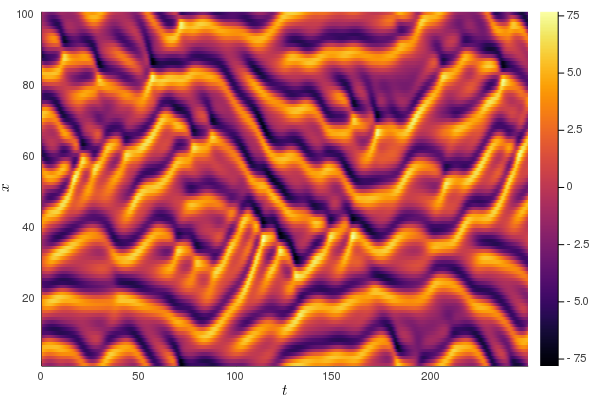}
    \caption{
    A space-time plot of a solution of the KS equation for $L=100$, $0<t<250$.
    (From \url{https://commons.wikimedia.org/wiki/File:Kuramoto–Sivashinsky_spatiotemporal_evolution.png}, provided with CC0 licence.)}
    \label{fig:KS}
\end{figure}

Like the Lorenz equation, the KS equation can describe a chaotic system. 
As such, it is a difficult system to predict. A formulation of the equation in one spatial dimension, which is used in a prediction task, is:
\begin{equation}
    y_t = - yy_x - y_{xx} - y_{xxxx}
    \label{eq:k-s-1}
\end{equation}
where $y_t = \partial y / \partial t$ and $y_x = \partial y / \partial x$.
The initial condition ($t=0$)
is usually defined on a finite $x$ domain ($0 \leq x < L)$ with periodic boundary conditions.
The time behaviour is periodic for small $L$;
as $L$ is increased, the periodic behaviour undergoes bifurcations,
and eventually becoming chaotic (Figure~\ref{fig:KS}).

The KS equation has been used in the context of Reservoir Computing. Initially it was used as an actual task \citep{Pathak2017-fj}: here the Reservoir Computer's task is to predict the next output of the equation, to test if the Reservoir Computer can imitate the Lyapunov exponents of the equation, whether or not divergence takes place. The KS equation is used to find out whether Reservoir Computing could solve the observer problem, which involves deducing the full state of a dynamical system based on partial information of the state and of the system dynamics \citep{Lu2017-ir}.

The KS equation has also been used as a prediction benchmark,
with spatial sizes of $L=200$ \cite{Vlachas2019-rq}
and $L \in \{12, 22, 36, 60, 100\}$ \cite{Mushegh_Rafayelyan_Jonathan_Dong_Yongqi_Tan_Florent_Krzakala_Sylvain_Gigan2020-lb}.

Pathak et al. \cite{Pathak2018-bq} also compare against a modified form, to investigate the effect of  spatial inhomogeneity:
\begin{equation}
    y_t = - yy_x - y_{xx} - y_{xxxx} + \mu \cos(2\pi x/\lambda)
    \label{eq:k-s-2}
\end{equation}
where the size $L$ is an integer multiple of the wavelength $\lambda$,
with $L \in \{100, 200, 400, 800, 1600\}$.

\subsection{Sunspot numbers: unknown dynamics}\label{sec:sunspots}
Sunspots are dark spots that appear on the sun on a temporary basis. Sunspots have been recorded consistently since 1610 \citep{Ngdc_undated-yi}. As such, the observations can be used as a dataset for the prediction of the behaviour of an unknown dynamical system. 

Predicting the next value in a dataset of sunspots is a perennial task in Machine Learning, first used in the early twentieth century \citep{Yule1927-ne}. The task varies greatly in the details: while each involves predicting the next value in the dataset, which dataset is used, how the values are calculated, and what preprocessing is applied, all vary across the literature. The only consistency appears to be that the datasets typically start in 1749, although there remains some inconsistency on whether the first value should be taken in January \citep{Stepney:2021-UCNC} or July \citep{Schwenker2009-og, Rodan2011-xb}.

\begin{table*}[tp]\small
\begin{center}
\begin{tabular}{@{}lcc@{}}
\toprule
 dataset name & source & used in\\
\midrule
  NOAA dataset & \cite{Ngdc_undated-yi} & \cite{Schwenker2009-og}\\
  \specialcell{Zurich Monthly sunspot numbers}& \cite{Brownlee2016-yg} & \cite{Stepney:2021-UCNC}\\
  \specialcell{NASA Greenwich sunspot numbers} & \cite{noauthor_undated-nu} & \cite{Dale2018-rr}\\
  \specialcell{Carrington sunspot numbers} & \cite{noauthor_undated-ci} & \cite{Shougat2021-wd}\\
\bottomrule
\end{tabular}
\caption{
Sources found in the literature for the sunspot numbers, and an example citation of the source's use in the literature as the `Sunspot Prediction' benchmark.}\label{table:sunspots}%
\end{center}
\end{table*}

\begin{figure*}
    \centering
    \includegraphics[width=\linewidth,clip,trim= 3cm 0 3cm 0]{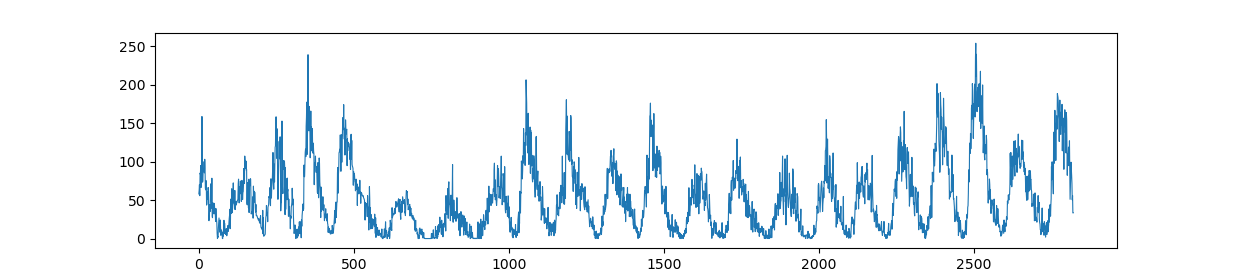}
    \caption{
    The Zurich Monthly Sunspots Numbers}
    \label{fig:zurich-sunspots}
\end{figure*}

At least four different sources for sunspot numbers are used in the literature, as shown in table~\ref{table:sunspots}; the Zurich dataset is shown in figure~\ref{fig:zurich-sunspots}. 
These have differing ways of dealing with missing data. 
In the NGDC dataset, a value of $-99$ indicated missing data;
other datasets appear to have removed this data in preprocessing. 
Some authors who use the NGDC dataset refer to preprocessing to deal with missing data, but do not define how this is performed \citep{Schwenker2009-og}. 

As well as the existence of differing datasets, other problems with this task are found. 
There is the inconsistency of the time period used: some use months \citep{Stepney:2021-UCNC}, others days \citep{Shougat2021-wd}. 
There does not  seem to be any specification on how monthly averages are computed, and whether or how they account for the different lengths of months. 
There is also the impossibility of making perfect observations:
issues may arise from cloud cover leading to missing data, quality of telescopes, or difficulty deciding on how to count sunspots.
All these render this data unreliable as data about a dynamical system, and make comparison between different systems problematic.

That said, finding out whether a data-driven model driven by imperfect data can produce usable results is a potentially valuable problem; all real world data is, after all, imperfect: data is merely a representation of the real state, and like all representation, some details are abstracted away. Having a dataset where some of the limitations are known and acknowledged thus has its own value; but if this is the case, it is important for users of the dataset to acknowledge this.

\subsection{Santa Fe Laser dataset: unknown dynamics}\label{sec:Santa-Fe}
The Santa Fe Laser dataset is a dataset originally produced in an experiment trying to replicate Lorenz-like chaos in NH3–FIR lasers \citep{Hubner1994}. It was initially distributed as one of the datasets in the Santa Fe 1992 Time Series prediction competition, the proceedings of which were published in 1994 \citep{Weigend2018}. The dataset is now frequently used as a prediction task for an unknown system.

\begin{figure*}[tp]
    \centering
    \includegraphics[width=\linewidth]{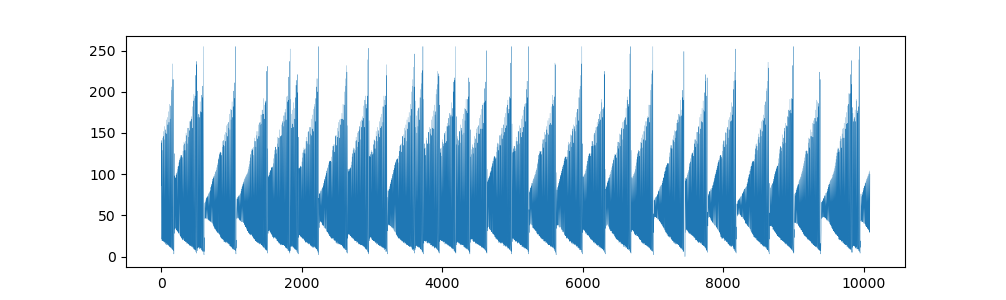}
    \caption{ The Santa Fe laser dataset.}
    \label{fig:santa-fe}

    \includegraphics[width=\linewidth]{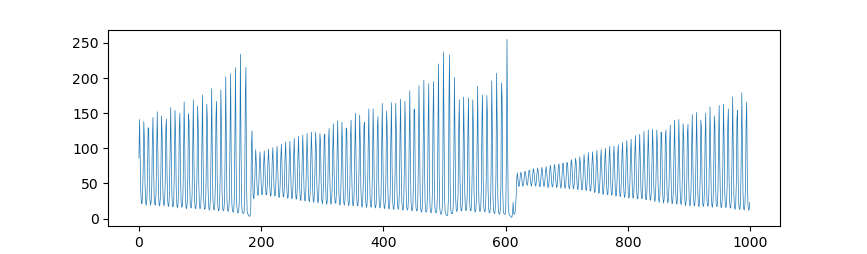}
    \caption{ Zoom in on the first 1000 values of the Santa Fe laser dataset.}
    \label{fig:santa-fe-short}
\end{figure*}

The original dataset was only 1000 data points long, the shortest of the datasets distributed in the competition. It was also noted for being stationary, low dimensional, clean, scalar-based, and nonlinear. A particularity of the dataset is that the data is characterised by catastrophes: the values grow in a somewhat predictable manner, until one of these catastrophes take place and the values change drastically. See figures \ref{fig:santa-fe} and \ref{fig:santa-fe-short}.

The most readily available version of the dataset, with the original 1000 values, can be found at \cite{Cph2021-ab}\footnote{the full dataset is available from: \url{https://github.com/MaterialMan/CHARC/blob/master/Support\%20files/other/Datasets/laser.txt}
}. 
The amount of data used varies in the literature, and is occasionally not recorded, as we can see in the examples listed in table~\ref{table:santa-fe}.

\begin{table*}[tp]\small
\begin{center}
\begin{tabular}{lccc}
\toprule
 source & total length & training & testing\\
\midrule
\cite{Larger2012-og} & not recorded & not recorded & not recorded\\
\cite{Brunner2013-hm} & 4000 & 3200 & 800\\
\cite{Dale2018-rr} & 5000 & 25000 & \specialcell{1250 for validating individual reservoirs,\\1250 more for evaluating  evolved reservoirs}\\
\cite{Guo2019-fl} & 4000 & 3000 & 1000 \\
\cite{Shougat2021-wd} & not recorded & not recorded & not recorded\\
\bottomrule
\end{tabular}
\caption{ The parameter values used in the literature for the Santa Fe Laser benchmark.}\label{table:santa-fe}
\end{center}
\end{table*}
The different experimental setups may lead to issues in interpreting results, as they may be unaccounted for independent variables; in such cases, it becomes difficult to distinguish where the variation in results comes from.

Some of this variation, such as the topology of the reservoir when this is randomly selected, may be smoothed out by repeating the experiment enough times. However, differing experimental setups may lead to systematic errors that persist over each experiment in a publication. 

The Santa Fe 1992 Time Series Prediction competition also distributed five other datasets. These are not used as benchmarks, except for the sleep apnea dataset (see section~\ref{sec:Santa-Fe-apnea}). 
It is generally safe to assume that any work referring to the `Santa Fe benchmark' 
with no further elaboration is using the laser dataset.

\subsection{McMaster IPIX radar data : unknown dynamics}\label{sec:ipix radar}
The McMaster IPIX Radar is an X-band radar, originally created to detect icebergs, but capable of collecting data on all sea clutter. 
The website~\citep{noauthor_undated-hq} indicates that the movement of sea clutter is influenced by several dynamical systems, and is itself a nonlinear dynamical system. As such, it makes an ideal prediction benchmark for Reservoir Computing.

The radar’s website makes two databases available for general use~\citep{noauthor_undated-hq}, one created from data recorded in Dartmouth, Nova Scotia, in 1993, and the other from data collected in Grimsby, Ontario, in 1998. The latter database is marked as incomplete by the radar’s website, and is encouraged to be used only comparatively with the Dartmouth data.

This data has been used as a prediction task in Reservoir Computing~\citep{Xue2007-df}.  It is unclear exactly which of the datasets are used, or how these are provided as input for the Reservoir Computer.  2000 datapoints are used, of which 200 are used as washout data, 800 for training, and 1000 for testing, but no more information is given. 

Subsequent uses of this dataset in Reservoir Computing \citep{Cernansky2008-fv,Rodan2010-ku,Rodan2011-xb, Duport2012-cw, Antoine_Dejonkheere_Francois_Duport_Anteo_Smerieri_Li_Fang_Jean-Louis_Oudar_Marc_Haelterman_Serge_Massar2014-dm} appear to be  consistent with the first use, citing \citep{Xue2007-df} and using the same dataset sizes.

\section{Computation tasks}\label{comp-tasks}

\subsection{XOR Task: binary output}\label{sec:xor}
The XOR task is a family of tasks based on the bitwise operation XOR, and entered the Reservoir Computing field through the field of Neural Networks. Unlike many of the benchmarks discussed here, there is no single XOR task, nor consensus on what the goal of the task is. Here we define it as a single-output classification task,
and describe some variants, to demonstrate the ways the task can be used.

An XOR task is used early in the history of Reservoir Computing,  where a bucket of water is shown to support a simple Liquid State Machine \citep{Fernando2003-oo}. The task is to output the XOR of the two simultaneously applied inputs.  Much post-processing is needed to read the output.
The task of outputting the XOR of two consecutively supplied input bits is also used \citep{Laporte2021-oe,Shougat2021-wd}.

The XOR task has been adapted to measure the memory of a Reservoir Computer as well as the nonlinearity \citep{Verstraeten2010-ns}: the task is to output the XOR of two initial values after a sizeable delay. A similar task has been used to benchmark the memory of an ESN \citep{Jaeger2012-ze}.

\subsection{Parity task: binary output}\label{sec:parity}
The parity task extends the XOR task to multiple inputs: its task is to distinguish an odd from even number of `high' inputs, to compute the parity. 

Here we designate this a computation task: it computes the parity.
It can also be considered a temporal classification task, or a known dynamical system imitation task, depending on one’s viewpoint.

Given a stream of inputs $u(t)$, the output of the PARITY-$n$ task is given by the parity of $n$ consecutive bits, read out after a delay $\tau$ \citep{Haykin2007-vx}:
\begin{equation}
    \label{eq:parity}
    y(t) = \mathit{PARITY}(u(t - \tau), u(t - \tau - 1), ... u(t - \tau - n))
\end{equation}

The input $u(t)$ is randomly sampled from $[-1, 1]$ \citep{Dion2018-xg} or $[0, 1]$\citep{Haykin2007-vx}.

The parity task has been used in reservoir computing, with a range of different $n$ values (see table \ref{table:parity}). 

\begin{table}[tp]\small
\begin{center}
\begin{tabular}{lc}
\toprule
 source & $n$\\
\midrule
 \cite{Haykin2007-vx} & 3\\
 \cite{Schrauwen2008-fx} & 5\\
 \cite{Dion2018-xg} & 2--5\\
 \cite{Shougat2021-wd} & 4, 6\\
\bottomrule
\end{tabular}
\caption{
Some of the different parameter values used for the RC parity benchmark.}\label{table:parity}%
\end{center}
\end{table}

\section{Classification tasks}\label{classification-tasks}
Classification tasks have categorical outputs (section~\ref{sec:class-cat-out})

\subsection{MNIST handwritten digits : static classification}\label{sec:mnist}
The handwritten digits benchmark is a benchmark commonly used to identify handwriting, and is thus a multiple-output classification task.

The MNIST Handwritten Digits Database \citep{noauthor_undated-lp} is a database of digits between 0 and 9 handwritten by high-school students and United States Census Bureau employees. The database consists of multiple datasets, most notable among which is the training dataset, composed of 60,000 images created by 250 writers. While originally black and white, the images were normalised to fit into $20 \times 20$ pixel arrays, and subsequently centred in a $28 \times 28$ pixel image, with the resulting images being in greyscale. 

This task has been used as a Reservoir Computing benchmark. \citep{Du2017-ch} describe a preprocessing technique that converts the image of the digit into a black-and-white $22 \times 20$ pixel image. 
Each row is then converted into a time series input fed into the reservoir. In order to make the task easier for the reservoir, it takes two inputs, sampled at different rates. It is unclear here how the data is fed in to the reservoir.

Another use of this database when benchmarking Reservoir Computing involves turning each image into a time series \citep{Schaetti2016-xx}, in this case feeding the image to the reservoir a single pixel at a time.  Another variation is to permute the pixels, in order to remove some of the internal structures of the image. This variation is called the Permuted Sequential MNIST task (psMNIST) \citep{Manneschi2021-vp, Manneschi2021-ir}.

\subsection{Spoken digits: non-stationary classification}\label{spoken-digits}
Identifying words that have been spoken is a classic task in neural networks, and is a multiple output category benchmark  used in Reservoir Computing almost since its inception. The goal of the task is to identify some words being spoken, whoever the speaker. The most common version of it uses the TI-46 dataset.

The TI-46 Isolated Spoken Words Dataset \citep{George_R_Doddington_and_Thomas_B_Schalk1981-ix} is a dataset of 20 individual words, spoken by 8 men and 8 women. It contains ten digits, and ten words common in speech recognition, such as `help' and `stop'. The dataset as available has 26 utterances per female speaker of each word digit, 10 for training and 16 for testing. The intent of the project was to have a dataset of words that personal assistant-style technology could be tested on, particularly ones suited to office environments, rather than home ones. The dataset can be found online \citep{noauthor_undated-fv}, but is not open access.

A subset of the TI-46 Dataset consisting of the digits from `zero' to `nine' said by five different speakers was first used in the context of Reservoir Computing to the speech recognition capabilities of Liquid State Machines to the then state-of-the-art Hidden Markov Models \citep{Verstraeten2005-fk}. This work was expanded to include ESNs \citep{Verstraeten2006-xa}.
The authors detail the experimental setup, which we summarise here:
\begin{itemize}
    \item Noise was added to the words, using the NOIS\-EX database  \cite{Varga1993-os}.
    \item The words were first preprocessed using the Lyon Passive Ear Model \citep{Lyon1982-ww}, a more biologically inspired model for preprocessing than the one performed for Hidden Markov Models. (Some other authors preprocess the data differently.)
    \item The words were then encoded into spike trains using the Bens Spiker Algorithm (BSA) \citep{Schrauwen2003-vf}.
    \item The evaluation was performed using the Word Error Rate (WER), calculated as  $(N_{\text{correct}}/N_{\text{total}})$
\end{itemize}
Verstraeten et al. \cite{Verstraeten2006-xa}  added a degree of confidence measure to the output, which increased the accuracy of their results. 

The Isolated Spoken Digits task is by far the most popular classification benchmark, particularly after its first use. 
See, for example, \cite{Verstraeten2007-wc, Butcher_undated-it, Appeltant2011-el, Duport2012-cw, Larger2012-og, Paquot2012-li, Brunner2013-hm,  Antoine_Dejonkheere_Francois_Duport_Anteo_Smerieri_Li_Fang_Jean-Louis_Oudar_Marc_Haelterman_Serge_Massar2014-dm, Soriano2015-oi, Vinckier2015-td, Dion2018-xg, Moon2019-du}.

 A simple version of this benchmark was used in one of the earliest reservoir computing publications  \citep{Fernando2003-oo}, getting a Reservoir Computer to recognise the spoken digits “0” and “1”, with these digits being input as recorded speech samples.
 
Isolated spoken digits used as a standard benchmark, as described here, is separate from the larger problem of general speech recognition, which attempts to identify words and phonemes from larger datasets \citep{F_Triefenbach_A_Jalal_B_Schrauwen_and_J-P_Martens2010-cq}.


\subsection{Japanese vowels : non-stationary classification}\label{sec:vowel}
Another speech recognition task uses a dataset\footnote{%
\cite{Jaeger2007-mb} gives links to the dataset; 
those links no longer work.
The dataset can currently be found at \url{https://github.com/MaterialMan/CHARC/tree/master/Support\%20files/other/Datasets/JapaneseVowels}.
} of vowels spoken by nine Japanese men \citep{Jaeger2007-mb}. Unlike the spoken digits task, the task is not to categorise by what has been spoken, but rather to categorise by speaker. Hence it remains a multiple-output classification task.

In the original paper, four experimental setups were tested, the most successful of which we summarise here.
The experiment used 9 output nodes, one for each speaker, and had a training length of 270 samples. 
The inputs were processed by first being partitioned into $D=3$ subsequences of equal length; these subsequences were then joined into single input vectors of size $D$. The outputs were then trained using linear regression. Because overfitting was a concern, the reservoir size was kept to $N=4$.

While this task is not frequently used, it is interesting to contrast its goals to the more popular Spoken Digits Recognition task.

\subsection{Santa Fe sleep apnea dataset: stationary classification}\label{sec:Santa-Fe-apnea}

  \begin{figure*}
      \includegraphics[width=\linewidth]{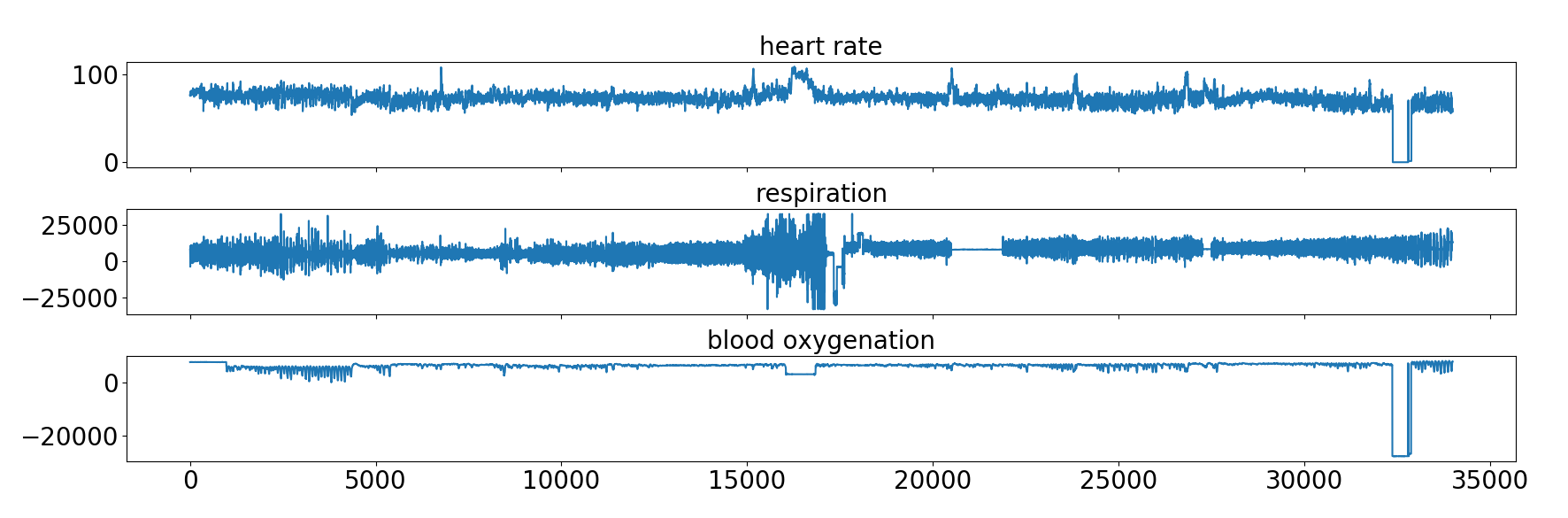}
      \caption{ Full sleep apnea dataset.}\label{fig:sleep-apnea-full}
  \end{figure*}

  \begin{figure*}
      \includegraphics[width=\linewidth]{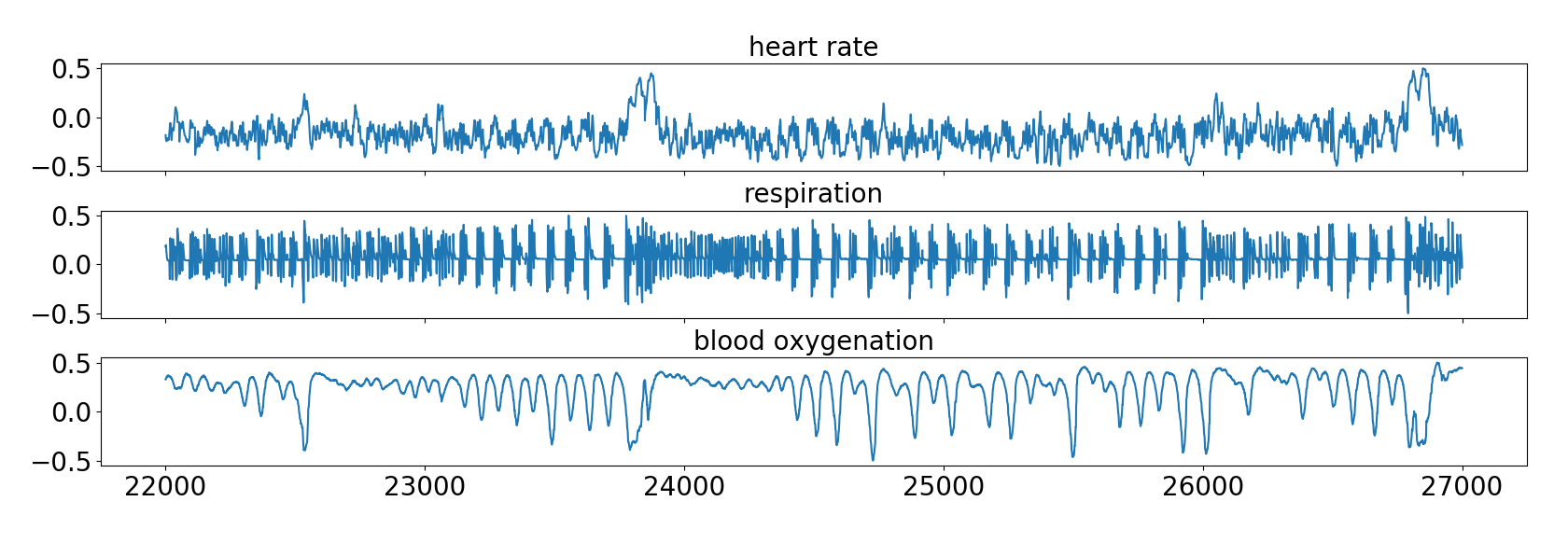}\caption{
      Portion of the sleep apnea dataset, as used in \cite{wringe2024ucnc}; each dataset portion is separately normalised to the range $[-0.5, 0.5]$ using equation \ref{eq:sleep-apnea-norm}.
      }\label{fig:sleep-apnea-normalised}
  \end{figure*}

The Santa Fe 1992 Time Series Prediction competition sleep apnea dataset \citep{Ichimaru1999-uz}, available at PhysioNet \citep{doi:10.1161/01.CIR.101.23.e215}
  comprises various physiological readings of a patient with suspected sleep apnea. The patient experienced both normal sleep and apnea periods, which can be seen in the data (figure~\ref{fig:sleep-apnea-full}).
  
  The data was gathered over 4 hours and 43 minutes with readings taken every 0.5 seconds, then digitised at 250 Hz. It is part of the Santa Fe Time Series Prediction Competition \citep{Schwenker2009-og},
  in two files, b1.txt and b2.txt, each with 17000 datapoints, with an accompanying file describing the content. 
  The dataset consists of readings of the patient's (i) heart rate, (ii) respiration rate, and (iii) blood oxygen saturation, although there is conflicting information as to whether the respiration data is derived from the patient's nasal airflow \citep{Schwenker2009-og} or chest volume \citep{Ichimaru1999-uz}.

  This benchmark is notable for having three related but distinct datasets (see figures \ref{fig:sleep-apnea-full} and \ref{fig:sleep-apnea-normalised}), as well as having a number of peculiarities which make it particularly challenging.

We designate this as a stationary temporal classification task,
as it could potentially be used to classify normal \textit{v} apnea periods.
However, in the literature, the dataset tends to be used for a time series prediction task, either predicting each datastream individually, or using a combination to improve prediction.
For example, \cite{Shougat2021-wd} treats each of the datasets as entirely distinct benchmark tasks, while \cite{wringe2024ucnc} studies how the datasets interact with each other.

Shougat et al. \cite{Shougat2021-wd} state that the input is normalised before being input to the reservoir, but it is unclear how this normalisation is performed. 
Wringe \cite{wringe2024ucnc} normalises the inputs to fall between $[-0.5, 0.5]$ using: 
  \begin{equation}
        {x'} = \frac{{x} - \text{min}({x})}{\text{range}({x})} - 0.5
        \label{eq:sleep-apnea-norm}
    \end{equation}

The data lengths used in \cite{Shougat2021-wd} are not stated. \cite{wringe2024ucnc} starts input at point 22000 (fig. \ref{fig:sleep-apnea-normalised}), and uses 1000 points as washout, 3000 points as training, and 1000 points as testing. These values are chosen to ensure both the training and testing phases include both normal sleep and apnea periods.

\section{Direct Property Measures}
\label{sec:property-measures}

\subsection{Memory Capacity}\label{sec:mc}
\subsubsection{Linear Memory Capacity}\label{sec:lmc}


Certain tasks are interesting because they directly reveal properties of a given Reservoir Computer. One of these is the Linear Memory Capacity task, which quantifies the fading memory of the reservoir. This was  introduced by Jaeger \cite{Jaeger2002-ml}, and has been investigated in the context of reservoir computing and used as a benchmark task  \citep{Verstraeten2007-wc, Rodan2010-ku, Duport2012-cw, Goudarzi2015-gv, Soriano2015-oi}.

The Linear Memory Capacity of a given Reservoir Computer that takes scalar input is defined as follows.
Consider an input stream $u(t)\in U[-1,1]$.
Train the reservoir to reproduce this input, delayed by a number of timesteps $k$:
the target output is ${\hat{v}} = u(t-k)$,
the observed output is $v_k(t)$, the linear memory capacity for this delay is defined as
the covariance squared of the delayed input (target output) and the observed output, 
normalised by the variances of the input and the observed output \cite[eqn.13]{Jaeger2002-ml}:
\begin{equation}
        MC_k = \frac{cov^2\left( u(t-k),v_k(t) \right)}{var(u(t)) var(v_k(t))} 
\end{equation}
The total Linear Memory Capacity is the sum over all delays \cite[eqn.15]{Jaeger2002-ml}:
\begin{equation}\label{eqn:MC}
        MC = \sum_{k=1}^{\infty} MC_k
\end{equation}
For higher $k$, each $MC_k$ tends to decrease (and the corresponding validation NRMSE tends to increase),
which means that inputs after longer delays are remembered less well, and are dominated by noise.
In the formal definition of $MC$ (eqn.~\ref{eqn:MC}), the sum over delays goes to infinity;
however the small values at high $k$ are essentially noise, and should be neglected,
so in practice a cutoff is used.
Dambre et al. \cite[SupMat3.2]{Dambre2012-lv} define a threshold based on the size of the reported capacity.
Dale \cite{Dale2018-rr} uses a cutoff of $k_{max} = 2N$ (number of nodes).

All the values of $MC_k$
can be found in a single run:
instead of training a target scalar output for a single $k$ using ${\hat{v}} = u(t-k)$,
train for all $k$ up to the relevant threshold,
using target vector output $\hat{\mathbf{v}}(t) = (u(t-1), u(t-2), \ldots, u(t-k_{\max}))^T$.

\subsubsection{Nonlinear Memory Capacity}\label{sec:IPC}

While the Linear Memory Capacity measure gives a useful quantification of a reservoir's memory capacity, it cannot on its own model the full computing power of dynamical systems.
\citep{Dambre2012-lv} generalise the definition of linear memory capacity in a way that allows them to define various non-linear capacities, too: they call this \textit{information processing capacity} (IPC).
These measure a Reservoir Computer's ability to compute a nonlinear function of past inputs, for example, a cubic function of delayed inputs, such as $u^3(t-1)$ or $u(t-1) u^2(t-2)$.
The cases where the polynomials involve different time delays are called cross memory capacities \citep{Duport2012:MC}.

While \cite{Duport2012:MC} considers quadratic polynomials,
\cite{Dambre2012-lv} considers complete sets of orthogonal polynomials, to cover all possible non-linear capacities and cross memory capacities with no double counting.
They use normalised Legendre polynomials for inputs $u(t)\in [-1,1]$.
They note that different  sets of orthogonal polynomials should be used for different input distributions, for example, Hermite polynomials for Gaussian-distributed inputs.
Orthogonal sets of trigonometric functions can also be used as the basis.

As for linear memory capacity,
the reservoir is trained to output the relevant (here, polynomial) function of its delayed input.  For a given degree $d$ of polynomial (linear, quadratic, cubic, etc), the contributions for all delays $k$, including all combinations of delays in the cross memory capacities, are summed.
Then the total non-linear memory capacity, or IPC, is the sum over all polynomial degrees.
The total IPC, from the contributions of all the linear and non-linear polynomials, is $N$ \cite{Dambre2012-lv}.

This process is computationally intensive, as there are many combinations of polynomials and delays as degree $d$ increases.
Contributions at high $d$ decrease, and a cutoff at large $d$ is used,
and the same observed state data can be used in all the training.

\subsubsection{NARMA as a memory capacity proxy}\label{sec:NARMA:IPC}

The NARMA benchmark (sec.\ref{subsubsec:narma}) also measures how well a reservoir can reconstruct a non-linear polynomial function of delayed inputs.
Despite the benchmark's limitations
(as discussed in sec.\ref{subsubsec:narma}),
the ability to successfully learn NARMA-$N$ can be used as a proxy for saying the reservoir has a memory capacity of $N$.

\subsection{Rank-based measures}
There are some task-independent measures that involve calculating the rank of a matrix constructed from multiple observations of the reservoir's state over time, when driven with random input.
The different rank measures depend on the form of the input and the state measurement points.

Two particular rank-based measures: kernel quality (kernel rank, KR),
and generalisation rank (GR), were first introduced in \cite{Legenstein-2007,Busing-2010}.
These have been adapted for use with ESNs in two different ways in the literature, discussed below.
Calculating the rank is discussed in section~\ref{sec:rank}.

\subsubsection{Kernel Rank}\label{sec:KR}

Kernel Rank (KR), also referred to as Kernel Quality, or Separation Rank, is a measure of how rich the nonlinear dynamics of the reservoir are. 
It measures how well the inputs are projected into a high dimensional state space, such that they can be separated by the linear output weight matrix.
High KR indicates good separability.

Vidamour et al. \cite{Vidamour2022} use a direct translation of the original definitions, and implement this measure as follows.
Consider $S$ maximally distinct input
streams, with values drawn from $U[-1,1]$, each of length $T$, and measure the reservoir state at the end of each input stream.
KR is the rank of the resulting $N \times S$ matrix $[\mathbf{x}_1 \mathbf{x}_2 \ldots \mathbf{x}_S]$, see algorithm~\ref{alg:KR Vidamour}.

\begin{algorithm}[tp]
\caption{Vidamour KR \cite{Vidamour2022}}\label{alg:KR Vidamour}
\begin{algorithmic}[1]
\State $S$ := number of input streams $\geq N$
\State $T$ := length of each input stream (timepoints)
\State run reservoir with washout stream
\For {$i \in 1..S$}
    \State $u_i(t), t\in 1..T$ := $U[-1,1]$ 
    \State run reservoir with input stream $u_i$
    \State $\mathbf{x}_i(T)$ :=  reservoir state at final time $T$ 
\EndFor
\State $\mathbf{X} := [\mathbf{x}_1(T)\, \mathbf{x}_2(T)\, \ldots\, \mathbf{x}_S(T) ]$ 
\State \Return   rank($\mathbf{X}$)
\end{algorithmic}
\end{algorithm}

Dale et al. \cite{Dale2019-rm} use a different adaptation, 
which is computationally less intensive (requiring $S$ inputs, rather than $S\times T$), and adapted to time series tasks (where the reservoir state at each timepoint is used), but is further from the original definitions.
In this approach, there is a single input stream, of length $S$, and the reservoir state is measured at each timepoint.
The stream's values are again drawn from $U[-1,1]$, thereby making them maximally distinct.
See algorithm~\ref{alg:KR Dale}.

\begin{algorithm}[tp]
\caption{Dale KR and GR \cite{Dale2019-rm}}\label{alg:KR Dale}
\begin{algorithmic}[1]
\State $S$ := length of input stream $\geq N$
\State $r$ := range of input, 1 for KR, 0.1 for GR
\State run reservoir with washout stream
\For {$t \in 1..S$}
    \State step reservoir with input $u(t) \in U[-r,r]$
    \State $\mathbf{x}(t)$ :=  reservoir state at time $t$ 
\EndFor
\State $\mathbf{X} := [\mathbf{x}(1)\, \mathbf{x}(2)\, \ldots\, \mathbf{x}(S) ]$ 
\State \Return  rank($\mathbf{X}$)
\end{algorithmic}
\end{algorithm}

Note that these two algorithms are the same in the case that $T=1$.

\subsubsection{Generalisation Rank}\label{sec:GR}
Generalisation rank (GR) measures how robust the
reservoir is to noise and avoiding overfitting.
The intent is to produce a measure of whether the matrix can generalise over inputs that are similar.
Low GR indicates good robustness to noise.

GR is computed in a similar manner to KR, but instead of running the reservoir over maximally different inputs, the reservoir is fed similar inputs, each with a small amount of noise added. 

Vidamour et al. \cite{Vidamour2022} again use a direct translation of the original definitions.
Consider $S$ input
streams, each of length $T$, each with values drawn from $U[-1,1]$, except that the last few values in each stream are set to be the same for all streams.
Measure the reservoir state at the end of each input stream.
GR is the rank of the resulting matrix, see algorithm~\ref{alg:GR Vidamour}.

\begin{algorithm}[tp]
\caption{Vidamour GR  \cite{Vidamour2022}}\label{alg:GR Vidamour}
\begin{algorithmic}[1]
\State $S$ := number of input streams $\geq N$
\State $T$ := length of each input stream (timepoints)
\State $\tau$ := length of common input stream 
\State $tail(t), t\in 1..\tau$ := $U[-1,1]$
\State run reservoir with washout stream
\For {$i \in 1..S$}
    \State $u_i(t), t\in 1..(T-\tau)$ := $U[-1,1]$ 
    \State $u_i$ := $u_i + tail$ \Comment append the tail
    \State run reservoir with input stream $u_i$
    \State $\mathbf{x}_i(T)$ :=  reservoir state at final time $T$ 
\EndFor
\State $\mathbf{X} := [\mathbf{x}_1(T)\, \mathbf{x}_2(T)\, \ldots\, \mathbf{x}_S(T) ]$ 
\State \Return  rank($\mathbf{X}$)
\end{algorithmic}
\end{algorithm}

Dale et al. \cite{Dale2019-rm} again use a different adaptation of the original definitions.
The algorithm for GR is the same as for KR,
except that the streams' values are drawn from a reduced range $U[-0.1,0.1]$, thereby making them similar.
See algorithm~\ref{alg:KR Dale}.


 \subsubsection{Calculating the rank}\label{sec:rank}

The standard way to calculate the rank of a matrix
is to use singular value decomposition (SVD).
The \textit{rank} of a matrix is the number of  non-zero singular values.
Rank is an integer, so, for small reservoir size $N$, this can result in a very granular measure.

Due to numerical effects and noise, typically \textit{all} the singular values are non-zero, but some may be very small.
To make the measure meaningful (rather than always $N$)
in practice a threshold is chosen, below which the singular values are taken to be effectively zero.
This threshold is typically expressed as some percentage of the maximum singular value.
The threshold value is essentially arbitrary,
and affects the measured rank,
so should be stated in any results.

An alternative to this integer-valued rank is the
 real-valued \textit{effective rank} \citep{Roy_undated-zm,Love2021-du}. 
Normalise the singular values $\sigma_i$ to sum to one: $p_i = \sigma_i / \sum_i \sigma_i$.
The effective rank is defined as $\exp(-\sum_i p_i \ln p_i)$.
If $R$ of the singular values are the same, and the rest are zero, then $p_i = 1/R$ or 0, giving an effective rank of $R$,
which is the same as the standard rank value.  For other cases, effective rank gives continuous values, has no arbitrary cutoff, and weights the singular values according to their size, potentially giving a more meaningful result.

The number of measured states $S$ should not be less than $N$, since the rank of the resulting matrix is  $\leq \min(S,N)$.
The measured rank increases with $S$ until it eventually converges \cite[sec.D.a]{Dale2019-rm}.
Preliminary investigation should be performed to establish a suitable value for $S$.

 \subsubsection{Generalising the rank measures}\label{sec:genrank}

Given the issues surrounding the definition and comparison of KR and GR in the context of reservoir computing,
other measures are being developed to capture similar information, while being better defined.
For example, a family of entropy transformation measures,
that evaluate how the entropies of the input and output data are related,
have been developed \cite{Griffin:2024-UCNC,Griffin2025}.

\subsection{Benchmarks for `free'}

Certain tasks, such as MC and GR, can be calculated from the same data output from the ESN \citep{Love2021-du}. Both tasks use the same input; the observations of the state can be used to calculate  the KR, and an output layer can be trained on the same observations to find the memory capacity. This is useful when using slow physical reservoirs, or computationally intensive search methods such as CHARC.

\subsection{CHARC}
Memory capacity, Kernel Rank, and Generalisation Rank are all relevant properties of Reservoir Computers, but the aim should not be to maximise them all. There is no agreed upon principle that allows us to correlate these values of these metric. Indeed, different tasks are best accomplished by reservoirs with different properties \citep{Dale2019-rm}. Given this, they construct a framework based on Linear Memory Capacity, Kernel Rank, and Generalisation Rank. The Reservoir can then be evaluated not through its performance at the tasks used to measure these properties, but instead by the breadth of the behaviour space available to the substrate or model it is built upon.

When using these property measures, including in the context of CHARC, 
care needs to be taken to ensure that any arbitrary constants are fully documented to allow comparison between different results.
These include washout times, input stream lengths and data ranges, and particular rank algorithm including thresholds.

Dale et al. \cite{Dale2019-rm} map four of the benchmark tasks described here to their positions in the CHARC behaviour space. One piece of interesting further work might be to map more of them, in order to enable studies to select tasks that not only take a range of different forms, but also correspond to different areas of Reservoir behaviour spaces.

\section{Best practices for benchmarking}\label{best practice}

Having looked at many of the benchmarks used in the field of Reservoir Computing individually, we now  describe some best practices for using them. 
These can be considered the minimum criteria for reporting results and ensuring reproducibility.  However, some of the discussions above demonstrate that meeting these criteria can be a more subtle and involved task than it may appear at first glance.

Many of these best practices relate to experimental setups.
One of the advantages of using a benchmark is the ability to compare the results over different Reservoir Computers. These comparisons can typically be done by statistical tests. 
However, as it is not the norm to share one’s complete experimental data output sets, it is essential that researchers be able to reproduce each other’s experiments. For that reason, the experimental protocol of benchmark tests should be described in detail. 

This section primarily concerns running and reporting benchmarks in reservoir computing. For more general introductions and tutorials on designing, training and using RCs, see  \cite{Cucchi2022,Lukosevicius2012,Stepney-2024-NACO}.

\subsection{Datasets and data parameters}
\subsubsection{Using existing datasets}

\paragraph*{State which data source is used.}
Certain uses of cited benchmarks do not always point to the same dataset. This may be because there are several sources for the same kind of data, as with the various sunspot datasets (table~\ref{table:sunspots}). It may also be that a single source has recorded several sets of data, as is the case with the IPIX Radar (section~\ref{sec:ipix radar}).

\paragraph*{State which subset of the data is used.} 
There may be one single data source, but  only a subset of the data is used. This is the case with the Santa Fe Laser readings (section~\ref{sec:Santa-Fe}), of which the most commonly available source contains over 10,000 data points,
and is often sub-setted. 

\paragraph*{State what parameter values are used, and why these were chosen.}
Some equation-based benchmarks, such as the NARMA (section~\ref{subsubsec:narma}) and Mackey--Glass (section~\ref{sec:Mackey-Glass}) systems, rely on specific parameter values. Different values may lead to different behaviour, as with the Mackey--Glass system, where a $\tau$ of 16.8 or more will lead to chaotic behaviour. 
Sometimes several different sets of values are used in the literature.

\subsubsection{Introducing new datasets}

\paragraph*{State the algorithm and parameter values used in the dataset generation.}
As noted in \cite{DBLP:journals/corr/abs-2110-05266}, if a benchmark is based on a differential equation and numerically integrated, then the values used in the integration, such as timestep and grid size, or the details of a more sophisticated algorithm, can change the detailed generated dataset. 

\paragraph*{Provide suitable metadata along with new datasets.}
If using a new experimental or generated data set,
it needs to be published  with the results,
and accompanied by metadata.
Gebru  et al. \cite{gebru2021datasheets} provide comprehensive
guidelines for what metadata should be provided,
covering issues including
who created the dataset and who funded them, 
how and when the data was collected or generated,
ethical concerns, 
the structure of the data,
sampling, noise and errors, labelling,
data cleaning and pre-processing, 
previous uses, distribution, and maintenance.
Gilpin \cite{DBLP:journals/corr/abs-2110-05266}'s
catalogue of 131 chaotic dynamics benchmark systems
follows the approach given there.

\subsection{Experimental method}


\paragraph*{State all the reservoir parameters used.}

The first aspect of experimental setup that needs to be reported are the parameter values of the Reservoir Computer used. One example of this being done well can be found in \cite{Jaeger_undated-ek}, where the setup for the Mackey--Glass benchmark is described as having 400 nodes, bias input, inserted noise, and output feedback. When reporting an experiment involving an ESN, some parameters that should be reported are:
\begin{itemize}
    \item the number of nodes in the Reservoir State
    \item the input bias, if any
    \item the leakage rate, if any
    \item the details of any noise added to the input
    \item the details of any output feedback
    \item the connectivity of the Reservoir State
    \item the network topology, if not random
    \item the weight matrix distribution, sparsity, and scaling, and how  generated
\end{itemize}
A guide to setting these parameters can be found in \cite{Lukosevicius2012}.

When reporting an experiment involving an \textit{in materio} RC experiment, the parameters will be device dependent.
They should be reported in similar detail.


\paragraph*{State which subsets of the data source are used for washout, training, and testing, and justify the values chosen.}

Reservoir Computing tasks are typically composed of an input sequence of scalar of vector values, fed sequentially into the reservoir. For benchmark testing, experimental setups divide the input into three categories: washout, training, and testing. Washout data is ignored in the evaluation, is usually at least the length of the reservoir’s fading memory, and is used to ensure that the reservoir is driven solely by the inputs and not influenced by its initial state (see also section~\ref{sec:washout}). The training set is used to train the reservoir output weights, which are evaluated against the testing set.

When these sets are of different lengths across different experiments, this may affect the performance of the reservoir.

It is important to record the length of the washout set, particularly for common dataset-based benchmarks, where different washout lengths mean that the Reservoir Computer is being tested and trained on different subsets of the dataset. 
The training length of the reservoir can have an effect on its performance: up to a certain training length, increasing the training will lead to more consistent results across experiments, leading to a smaller standard deviation of different performances. After this saturation is reached, increasing the training length has little effect, and may result in overfitting. This saturation can be seen both in performance-based tasks such as NARMA, and value-based tasks such as Kernel Rank \citep{Dale2018-rr}.

The saturation length of a system will depend on properties of the reservoir, such as the size, as well as the difficulty of the task: therefore, there is no one training length that is optimal for all reservoirs and all tasks: researchers should instead find the saturation length and use that to benchmark. Reporting the saturation length for specific tasks and reservoir sizes may also give researchers another axis along which to compare the behaviour of different types of reservoirs. 

Similarly, different testing lengths may affect overall performance.
For example, in a free-running prediction task (section~\ref{sec:prediction}), the experiment that uses a longer testing set may yield worse performance, due to the accumulation of errors leading to growing errors  \citep{Jaeger2007-mb}.


\paragraph*{Gather data from multiple runs.} 

It is standard to perform multiple runs of a benchmarking experiment,
and get a range of results.
Any measure of performance that is taken over multiple runs with different inputs and/or internal weight values is more likely to measure how well the dynamics are emulated in general, as opposed to the ability to follow a specific sequence. 

The number of runs should be reported,
along with how the setup is changed for each run:
different input data, and/or different reservoirs (different random weight matrices for ESNs, different input matrices and physical configurations for \textit{in materio} reservoirs).

For datasets generated from known dynamics,
multiple runs on different inputs can be performed with different generated sequences, either generated from different random inputs, or from different time slices of the single generated stream.

With experimental datasets,
if the experimental data set is large enough, multiple training and testing sets can be sliced from the overall dataset.
In this case, the testing dataset may not immediately follow the training set, and so a separate testing washout period will be needed.
If the data set is not large enough for this approach,
multiple runs can still be performed on different reservoir configurations.

\paragraph*{State all other relevant experimental parameter values and algorithms.}

Various training parameter values may be used.
These should be reported in enough detail that the experiment can be reproduced.
For complex experimental setups,
a good way to clarify the method
is to use a pseudocode description of the experiments,
possibly abstracted from the experimental harness code.

\subsection{Evaluation Measures}

\subsubsection{Time-series evaluation measures}\label{sec:NRMSE}

With a few exceptions, dynamical systems-based benchmarks typically output a scalar value, rather than a multidimensional vector value, each timestep. 
Evaluation is then typically performed using one of Mean Squared Error (MSE), Root Mean Squared Error (RMSE), Normalised Mean Squared Error (NMSE), or Normalised Root Mean Squared Error (NRMSE).
However, other measures may also be used, such as
 Mean Absolute Error (MAE) and Mean Absolute Percentage Error (MAPE).

Given a desired target output time series $\mathbf{\hat{v}}=\{{\hat{v}}_t\}_{t=1}^{N}$, an observed output time series $\mathbf{v} = \{{v}_t\}_{t=1}^N$, with the mean $\langle \mathbf{x} \rangle \triangleq~ \frac{1}{N}\sum_{t=1}^N x_t$, these are defined by:
\begin{align}
    \mathit{MSE}(\mathbf{\hat{v}}, \mathbf{v}) 
    &\triangleq \frac{1}{N}\sum_{t=1}^N ({\hat{v}}_t - {v}_t)^2  
    \\ \nonumber
    &= \langle (\mathbf{\hat{v}} - \mathbf{v})^2 \rangle 
    \\
    \mathit{RMSE}(\mathbf{\hat{v}}, \mathbf{v}) &\triangleq \sqrt{\mathit{MSE}(\mathbf{\hat{v}}, \mathbf{v})}
    \\
    \mathit{NMSE}(\mathbf{\hat{v}}, \mathbf{v}) &\triangleq \frac{\mathit{MSE}(\mathbf{\hat{v}}, \mathbf{v})}{\mathit{MSE}(\mathbf{\hat{v}}, \langle \mathbf{\hat{v}}\rangle)}
    \\
    \mathit{NRMSE}(\mathbf{\hat{v}}, \mathbf{v}) &\triangleq \sqrt{\mathit{NMSE}(\mathbf{\hat{v}}, \mathbf{v})}
    \\
    \mathit{MAE}(\mathbf{\hat{v}}, \mathbf{v}) 
    &\triangleq \frac{1}{N}\sum_{t=1}^N |{\hat{v}}_t - {v}_t|  
    \\
    \mathit{MAPE}(\mathbf{\hat{v}}, \mathbf{v}) 
    &\triangleq \frac{100}{N}\sum_{t=1}^N \left|\frac{{\hat{v}}_t - {v}_t}{{\hat{v}}_t}\right|
\end{align}
A smaller error indicates better performance.
Normalisation makes the NMSE, NRMSE and MAPE measures dimensionless and independent of any scaling or units of the outputs,
so are more comparable across experiments with different systems.

Note that here normalisation is performed with respect to the target output,
\textit{not} the observed output \citep[p.661]{Lukosevicius2012}.
This makes baselining a given system's error using a constant  observed value well-defined.
In particular, the normalised measures NMSE, NRMSE, and MAPE are equal to 1 if each $v_t$ is set to the mean of the target values $\langle \hat{\mathbf{v}}\rangle$. 
Experimental values greater than one indicate very poor performance;
values less than one may also be achievable with just simple approaches (section~\ref{sec:base}).

Depending on what a given experiment is intended to convey, different choices of measure may be appropriate. 
To see how the difference between the measures, their behaviours are listed in table \ref{table:error-measures}. 
This shows how normalising the error removes the effect of changing the standard deviation, 
which itself varies with the choice of units or scaling.

\begin{table}[tp]
\begin{center}
\begin{tabular}{lcccc}
\toprule
std dev & 0.1 & 0.5 & 1 & 2\\ 
\midrule
MSE 	& 0.010 & 0.269 & 0.910 & 3.610 \\
RMSE 	& 0.101 & 0.519 & 0.954 & 1.900 \\
NMSE 	& 1.004 & 1.006 & 1.004 & 1.045 \\
NRMSE 	& 1.002 & 1.003 & 1.002 & 1.022 \\
MAE 	& 0.081 & 0.440 & 0.745 & 1.533 \\
MAPE 	& 100\% & 100\% & 100\% & 100\% \\
\bottomrule
\end{tabular}
\caption{
Various error measures for 100 randomly generated target values $\hat{v}$ drawn from a Gaussian distribution with a mean of 0 and different standard deviations, compared to an observed constant output value $v=0$. When the standard deviation of the target output varies, there is a marked effect on the error in the unnormalised measures, but the normalisation removes this difference.}\label{table:error-measures}
\end{center}
\end{table}

While these measures are appropriate for judging the performance of a task where there is an error margin, it is frequently unclear what has led works in the literature to choose one over the others. When benchmarking using an existing task, it is advisable to use the method existing in the literature, to enable direct comparisons. When creating a new task, however, it is worth considering different methods and choosing the one best suited to the task. 
The value of a normalised measure is more readily interpretable,
and the NRMSE is interpretable as a form of standard deviation.

\subsubsection{Classification evaluation measures}

There are many standard machine learning measures for classification success.
We mention just a few common ones here.
An appropriate choice for the particular benchmark experiment should be made, and documented.

For a binary classification, success (true positive and true negatives), false positives (type I errors) and false negatives (type II errors) are the simplest measures.
If there is some threshold or other classifier parameter,
a ROC (receiver operating characteristic) curve 
can show the performance as a function of that parameter value:
a larger area under the curve (AUC) indicates a better classifier.

For classification into a larger number of categories,
symbol error rate (the fraction of incorrect symbols  obtained) \citep{Jaeger2004-eu} may be used.
A confusion matrix plot of predicted versus actual category 
shows both the proportion of correct predictions, and the distribution of incorrect predictions.

\subsection{Presenting the results}

Present results as clearly as possible, with tables and pseudocode preferable to reporting it textually.

\label{sec:base}

\paragraph*{Calculate baseline success measures.}

As stated above, NRMSE = 1 can be achieved from a constant output set equal to the mean target output.
Hence an NRMSE less than 1 is typically claimed to be a success. However, naive predictions can easily result in NRMSEs below 1 for some benchmarks.
Thus a stricter success criterion is needed in these cases.

In time series based tasks, for example, a naive prediction is to use the previous target value, by setting $v(t) = \hat{v}({t-1})$.\footnote{%
This is the persistence model of weather forecasting: tomorrow will be like today \citep{AMS}.
}
To illustrate this, we calculate the NRMSE of this base case for three popular benchmark tasks (table \ref{table:base-cases}).
This demonstrates that the baseline for success should be NMRSE $\lesssim 0.8$ for NARMA-10, and $\lesssim 0.4$ for Zurich sunspots.

\begin{table*}[tp]
\begin{center}
\begin{tabular}{lc}
\toprule
  & $v(t) = \hat{v}(t-1)$ 
  \\
\midrule
NARMA-10 & mean=0.826, sd=0.044 
\\
Sunspots & 0.396 
\\
Santa-Fe laser & 0.969 
\\
\bottomrule
\end{tabular}
\end{center}
\caption[base case NRMSE]{\small
NRMSE of `base case' solutions for some common benchmark tasks: (i) NARMA-10, using parameters from table~\ref{table:narma-equations}, averaged over 100 runs each of 1000 data points;
(ii) Sunspots, using the full Zurich dataset;
(iii) the Santa-Fe laser task, using the full dataset.
}\label{table:base-cases}%

\end{table*}

\paragraph*{Perform and report statistical tests.}

When presenting results averaged over multiple runs,
present (as a minimum) means \textit{and} standard deviations (or medians \textit{and} quartiles)
to demonstrate both average behaviour, and how much variation is present in that behaviour.

When comparing results, perform the appropriate statistical tests for statistical significance and effect size.
If doing multiple comparisons, use a Bonferroni correction
to reduce the likelihood of false positives.

For more complex experiments, more sophisticated statistical tests may be appropriate.
See any standard textbook on statistics for definitions and choices of the appropriate tests.

\section{Potential pitfalls of consistency}

Consistency is important to be able to reproduce experiments or compare one’s results to that of others. However, if all authors test their Reservoir Computers on the same benchmarks using the same data, we encounter a problem: are Reservoir Computers good at capturing the dynamics of complex dynamical systems based on incomplete data, or are they simply good at predicting this specific sequence of values that we have chosen to represent  data? One must balance the ability to compare results with other tasks in the literature with the need to avoid overfitting. 

We propose that the way to strike this balance is by making any variation deliberate and documented, rather than accidental and based on missing information. Some of the possible approaches to this are detailed below; other approaches are also possible.

\paragraph*{Use a diverse set of benchmarks.} Different benchmarks require different behaviour from a Reservoir Computer \citep{Dale2019-rm}. One can choose benchmarks that span this behaviour space, as well as benchmarks that have different approaches to data and what should be done with it. 

\paragraph*{Use a mix of standard and bespoke benchmarks.} Certain authors choose to mix benchmarks that are common in the literature with ones introduced to illustrate something specific features of their work.
This allows them both to place their work within the literature, and to compensate for where those benchmarks may be lacking.

\section{Conclusion}
Benchmarks are a useful way of evaluating Reservoir Computing in different contexts, particularly in terms of performance, or perhaps `usefulness'. 
Many of these benchmarks are inherited from the wider Machine Learning community, while others have been developed specifically for the field of Reservoir Computing. This provides a varied pool to choose from, and allows authors to choose benchmarks based on the argument they wish to present, while still placing their use of benchmarks in context with other works in the literature. 

However, this mixture of sources has also led to a  less well-defined benchmarking culture than in many other fields: there is no one `Reservoir Computing Benchmark Suite', and even the closest equivalent, the NARMA benchmarks, has half a dozen different parameter sets and experimental setups that are inconsistent across different publications. This makes direct comparison between results of different works within the literature difficult, if not impossible.

New approaches such as CHARC \citep{Dale2019-rm} would allow us to survey a given substrate or reservoir model’s behaviour space, telling us whether a given substance would make a viable reservoir. While this is valuable work, however, with no direct link to benchmark tasks, it remains merely abstract. There is currently no direct mapping from benchmark tasks to areas within the behaviour space, and it is unclear if such a mapping would be possible. 

While this review is by no means comprehensive, we hope that it is a step towards bridging the gap between the various uses of benchmarks, and that it may help others make judicious choices in future work. 

\section*{Declarations}

\subsubsection*{Ethical Approval} 
Not applicable: no human or animal research involved.

\subsubsection*{Competing Interests} 

The authors report there are no competing interests to declare.

\subsubsection*{Funding} 
C.W. acknowledges support of a PhD studentship from the Department of Computer Science at the University of York.
M.T. and S.S. acknowledge partial funding from the MARCH project, EPSRC grant number EP/V006029/1.


\end{document}